\documentclass[twocolumn, pre, showpacs,amssymb,superscriptaddress]{revtex4-1}

\usepackage{epsfig, color}
\usepackage{amsthm, amsmath, amsfonts, ae, psfrag}
\usepackage{braket}
\usepackage{bbm}
\usepackage[english]{babel}
\usepackage{epstopdf} 
\usepackage{verbatim} 

\usepackage[colorlinks,
pdfpagelabels,
pdfstartview = FitH,
bookmarksopen = true,
bookmarksnumbered = true,
linkcolor = blue,
plainpages = false,
hypertexnames = false,
citecolor = blue] {hyperref}

\begin{document}

\title{Bulk structural informations from density functionals for patchy particles}

\author{Daniel~Stopper}
\email{daniel.stopper@uni-tuebingen.de}

\affiliation{Institute for Theoretical Physics, University of T\"ubingen, Auf der Morgenstelle 14, 72076 T\"ubingen, Germany}

\author{Frank Hirschmann}

\author{Martin Oettel}
\email{martin.oettel@uni-tuebingen.de}

\affiliation{Institute for Applied Physics, University of T\"ubingen, Auf der Morgenstelle 16, 72076 T\"ubingen, Germany}

\author{Roland~Roth}

\affiliation{Institute for Theoretical Physics, University of T\"ubingen, Auf der Morgenstelle 14, 72076 T\"ubingen, Germany}

\date{\today}

\begin{abstract}
We investigate bulk structural properties of tetravalent associating particles within the framework of classical density functional theory, building upon Wertheim's thermodynamic perturbation theory. To this end, we calculate density profiles within an effective test-particle geometry and compare to radial distribution functions obtained from computer simulations. We demonstrate that a modified version of the functional proposed by Yu and Wu  [J. Chem. Phys. \textbf{116}, 7094 (2002)] based on fundamental measure theory for hard spheres produces accurate results, although the functional does not satisfy the exactly known low-density limit.  In addition, at low temperatures where particles start to form an amorphous tetrahedral network, quantitative differences between simulations and theory emerge due to the absence of geometrical informations regarding the patch arrangement in the latter. Indeed, here we find that the theory fits better to simulations of the floating-bond model [J. Chem. Phys. \textbf{127}, 174501 (2007)], which exhibits a weaker tetrahedral order due to more flexible bonds between particles. We also demonstrate that another common density functional approach by Segura \textit{et al.} [Mol. Phys. \textbf{90}, 759 (1997)] fails to capture fundamental structural properties. 
\end{abstract}

\maketitle
 \section{Introduction} \label{Sec:Intro}
A valuable and intuitive model system for directional interactions is formed by the framework of patchy particles \cite{KernFrenkel2003, Glotzer2004, pawar2010fabrication}. In addition to a repulsive core, two particles interact via a specific amount of attractive bonding sites distributed on their surface. 
Compared to fluids interacting via purely isotropic forces, genuinely novel phenomena have been found, such as
empty liquids \cite{BianchiLargo2006, delasHeras2011_Emptyliquids}, reentrant phase behavior \cite{RussoTavares2011_PRL}, 
and equilibrium gel states more stable than crystals \cite{Smallenburg2013, SciortinoZaccarelli2017}. 
This appears to be relevant for experiments since there are many systems in soft-matter physics in which the interactions of the constituents is of patchy nature, e.g., DNA-nanoassemblies \cite{Biffi2013, Rovigatti2014}, colloidal clays \cite{Ruzicka2011} or 
proteins \cite{Sear1999, Sciortino2007_patchyColloidsProteins, Whitelam2010, RoosenRunge2014, Soraruf2014}. Besides these, directional interactions are present in many molecular fluids, the probably most prominent representatives being hydrogen bonds in water \cite{Andersen1973, Nezbeda1989, Ghongasi1993, Jackson2006}.

For bulk fluids of patchy (or associating) particles, Wertheim developed a successful thermodynamic perturbation theory \cite{WertheimI1984, WertheimII1984, WertheimIII1986, WertheimIV1986, Jackson1988}. In its first order formulation (TPT1), correlations between bonding sites are neglected, and loop-forming structures are not considered. Moreover, it is assumed that each site is engaged in not more than one bond, and each pair of particles can be connected by at most one bond. Finally, the theory includes no information about the geometrical arrangement of the patches on the particles surface. Importantly, Wertheim's approach has proven to be in good accordance with simulations in terms of bonding statistics and phase diagrams \cite{Jackson1988, Bianchi2008, Sciortino2011, Teixeira201716_review}. 

However, the theoretical description of inhomogeneous systems of patchy particles appears to be more challenging. In principle, Wertheim's theory can also be extended to inhomogeneous fluids \cite{KierlikRosinberg1992, Marshall2013} resulting in a density functional \cite{Evans1979} describing the association between particles. For one- and two-patch colloids this formalism has also been extended beyond TPT1 and yields good agreement with simulations at planar interfaces \cite{Marshall2013, Haghmoradi2016}. However, for tetravalent patchy particles (i.e., particles carrying four bonding sites) this approach gave relatively poor results compared to simulations \cite{Segura1997}. Instead, the perhaps most common method used in the literature is to employ the bulk form of TPT1 as a starting point: There is the DFT version of Segura \textit{et al.} \cite{Segura1997}, which makes use of the weighted densities of Tarazona (WDA) \cite{Tarazona1990} for both the hard sphere and association contribution. Another version is a density functional suggested by Yu and Wu \cite{YuWu2002} which employs the framework of Rosenfeld's fundamental measure theory \cite{Rosenfeld1989, Roth2010} (FMT) for hard spheres  in combination with TPT1. In particular, it phenomenologically incorporates the vector-type weighted densities of FMT. Both functionals were initially tested close to a hard, planar wall for particles carrying four patches. Here, good, even quantitative agreement, was observed over a wide range of densities and temperatures in comparison with computer simulations, with the FMT-based functional performing slightly better at higher particle densities and being less computationally intensive for mixtures. Deviations between WDA and FMT were fairly small. To date the Yu-Wu approach seems to have become \lq state-of-the-art\rq\, and commonly is applied, e.g., to investigate properties of associating fluids at planar \cite{GnandelasHeras2012, OleksyTeixeira2015, Fries2017} or spherical interfaces \cite{Bymaster2007, Hughes2013, Wang2017}. However, in situations in which the orientation of the particles becomes important, such as for trivalent patchy spheres close to a neutral wall, both 
density functional models break down \cite{GnandelasHeras2012}. This breakdown is a consequence of the facts that the density functionals do not depend on the particle orientations as well as the absence of informations regarding the geometric arrangement of bonding sites. 

Surprisingly, systematic studies addressing the bulk structural behavior of density functionals describing patchy particles based on TPT1 seem to be lacking. This is probably due to the fact that it is a priori not clear how to proper model the directional interactions between a test particle and its surrounding fluid within Percus' test particle approach \cite{HansenMcDonald2013} to obtain the (orientationally averaged) radial distribution function $g(r)$ of the fluid. In this work, we demonstrate that density profiles obtained from DFT around a test particle interacting via a short-ranged spherical symmetric square-well potential with its patchy neighbors, 
can excellently match simulation results of both $g(r)$ and the corresponding static structure factor $S(k)$. 
While in simulations this approach yields distinct results compared to $g(r)$ due to the fact that orientations of the particles in vicinity to the tracer are distinguished, we propose that it is precisely the design of the investigated density functionals which here leads to a very good agreement between theory and simulations in terms of the radial distribution function.

We test three different density functionals: (i) the WDA-based approach of Segura \textit{et al.} \cite{Segura1997}, (ii) the original FMT-functional by Yu and Wu \cite{YuWu2002}, and (iii) a slightly modified version (which we will refer to as mFMT) of the latter incorporating the vector-type weighted densities in a different manner. Considering two isochores, we find that only density profiles obtained from the mFMT functional match well to computer simulations over a wide range of temperatures. 
The FMT functional underestimates correlations in the fluid, and the WDA approach completely fails to capture fundamental structural aspects seen in simulations. At lower temperatures where an amorphous tetrahedral network starts to form, the mFMT also underestimates correlations compared to simulations. 
However, most likely this may be attributed to approximations introduced by Wertheim's perturbation 
theory: foremost, the missing spatial correlations between bonding sites. This is supported by simulation results of the 
so-called floating-bond model \cite{Zaccarelli2007, Bleibel2018}. The latter generates directional interactions via floating bonds that interact via a non-additive isotropic attractive potential with larger particles; interactions among the latter and floating bonds themselves are thereby purely repulsive. 
By employing suitable geometric constraints on the bond-particle interaction, it is possible to 
restrict the maximum number of bonds attaching to the surface of a larger particle, and to generate a system favoring tetrahedral ordering of the large particles at low temperatures. By means of a suitable temperature rescaling, we find 
surprisingly good agreement between the Wertheim theory and floating bonds. 
We suggest that this may be understood due to conceptual similarities between both model systems. 
 
 The paper is structured as follows. In Sec. \ref{Sec:ModelTheory} we first introduce the model for patchy particles (Sec. \ref{SubSec:PatchyParticles}) that is also used in computer simulations. This is followed by a brief overview on Wertheim's theory (Sec. \ref{SubSec:TPT1}). Section \ref{SubSec:DFT} provides an overview of density functional theory and density functionals employed in this work.  In Sec. \ref{SubSec:TestParticleApproach} we introduce the test particle approach, and subsequently, in Sec. \ref{SubSec:FloatingBonds} we discuss the floating-bond model as an alternative for modeling directional interactions. A comparison between theoretical results and simulations for bulk structural properties is presented in Sec. \ref{Sec:Results}. This is followed by a discussion in Sec. \ref{Sec:Discussion} and closed by a summary and outlook in Sec. \ref{Sec:Summary}.

\section{Models and Theory} \label{Sec:ModelTheory}

\subsection{Kern-Frenkel patchy particles} \label{SubSec:PatchyParticles}

  \begin{figure}[t] 
  	\centering
  	\includegraphics[width = 8cm]{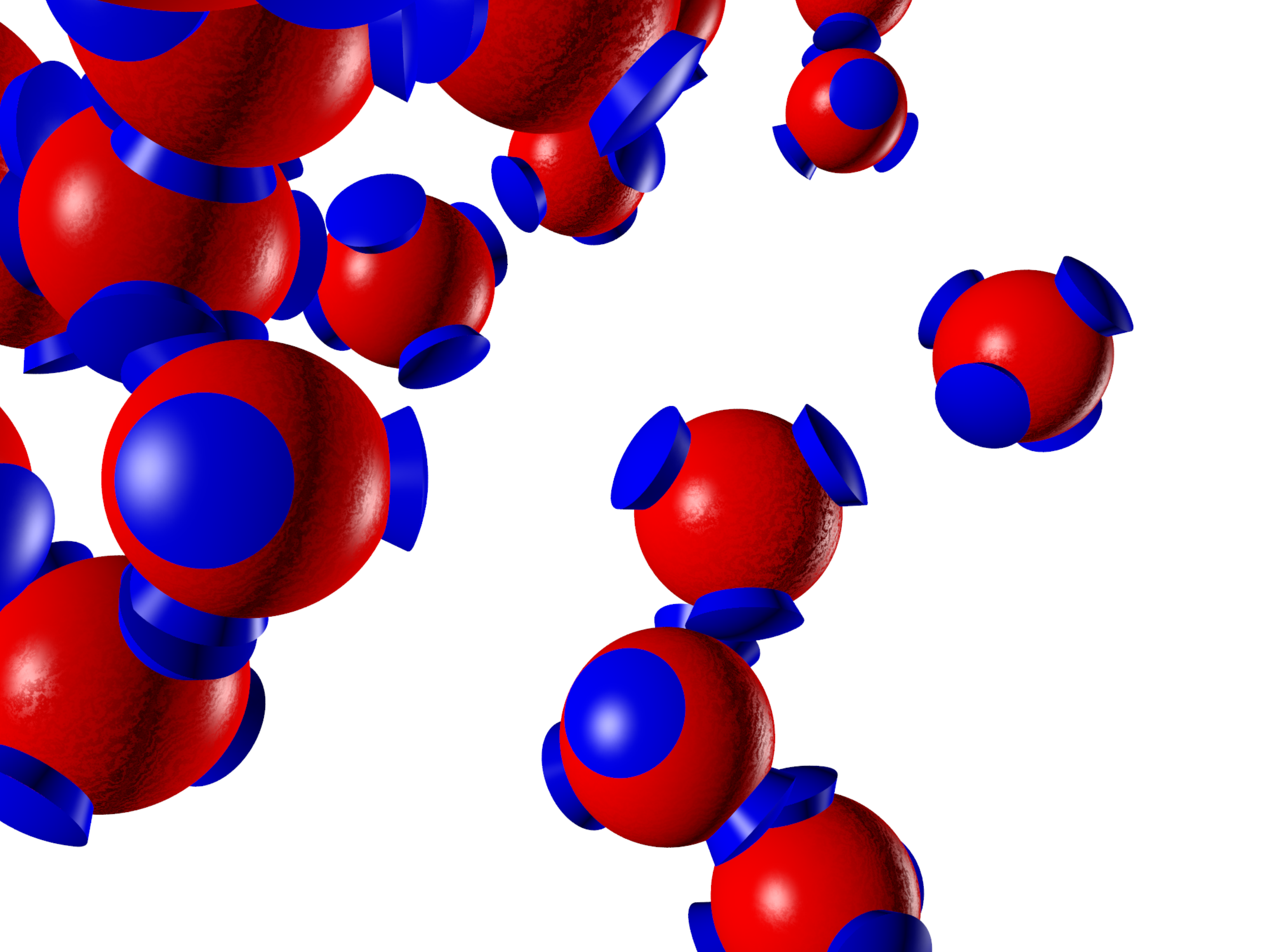}
  	\caption{Sketch of patchy particles with four patches (blue cones) interacting via the Kern-Frenkel potential \eqref{EqInteractionPotential}. When two patches of distinct particles overlap a bond is formed.} \label{Fig:snapshotKF}
  \end{figure} 
 
 We consider a system consisting of hard spheres with diameter $\sigma$, supplemented by a specific number $M$ of identical attractive bonding sites distributed in a regular pattern on their surfaces. This type of particles may be described by the Kern-Frenkel (KF) pair potential \cite{KernFrenkel2003} 
  \begin{align} \label{EqInteractionPotential}
  \phi(\mathbf{r}, \Omega_1, \Omega_2) &= \phi_\text{hs}(r) + \phi_\text{bond}(\mathbf{r}, \Omega_1, \Omega_2)
  \end{align}
  where $\phi_\text{hs}(r)$ is the usual hard-sphere potential, and $r = |\mathbf{r}| \equiv |\mathbf{r}_1 - \mathbf{r}_2|$ is the center-to-center distance between particles $1$ and $2$, and $\Omega_i$ denote individual orientation vectors of particles with label $i$.
  More specifically, the bonding interaction $\phi_\text{bond}(\mathbf{r}, \Omega_1, \Omega_2)$ is given by
  \begin{equation} \label{Eq:InteractionPotentialBonds}
  \phi_\text{bond}(\mathbf{r}, \Omega_1, \Omega_2) = \sum_{\alpha,\beta=1}^{M} \phi_\text{patch}(\mathbf{r} ,\hat{\mathbf{r}}_1^\alpha(\Omega_1), \hat{\mathbf{r}}_2^\beta(\Omega_2))
  \end{equation}
  where $\phi_\text{patch}$ describes the interaction between two patches $\alpha$ and $\beta$ of particle $1$ and particle $2$, respectively. The quantity $\phi_\text{patch}$ can be decomposed into a product of a spherically-symmetric square-well potential $\phi_\text{sw}(r)$ and a function $G(\hat{\mathbf{r}} ,\hat{\mathbf{r}}_1^\alpha(\Omega_1), \hat{\mathbf{r}}_2^\beta(\Omega_2)) ,\,$ containing the orientational character of the interaction. The square-well potential is given by
  \begin{equation} \label{SqSWInteraction}
  \phi_\text{sw}(r) = 
  \begin{cases}
  -\varepsilon~~~~~\text{if} ~~\sigma < r < \sigma+\delta\\
  0~~~~~~~~~\text{otherwise}\,,
  \end{cases}
  \end{equation}
  where the parameter $\delta$ controls the range of the attraction, and $\varepsilon$ is the potential depth. Both quantities are assumed to be identical for all patches. The square-well interaction introduces an unique energy scale in the system with a reduced temperature $T^* = k_B T / \varepsilon$. The orientational character of the patch-patch interaction may be written as \cite{KernFrenkel2003}
  \begin{equation}
  G(\hat{\mathbf{r}}, \hat{\mathbf{r}}_1^\alpha(\Omega_1), \hat{\mathbf{r}}_2^\beta(\Omega_2)) = 
  \begin{cases}
  1 ~~~~\text{if}~~ \begin{cases}
  \hat{\mathbf{r}}\cdot\hat{\mathbf{r}}_1^\alpha > \cos(\theta_c)\,,\\
  -\hat{\mathbf{r}}\cdot\hat{\mathbf{r}}_2^\beta > \cos(\theta_c)\,,
  \end{cases} \\
  0 ~~~~\text{else}~~\,,
  \end{cases}
  \end{equation}
  where $\hat{\mathbf{r}} = (\mathbf{r}_{1} - \mathbf{r}_2)/|\mathbf{r}_1 - \mathbf{r}_2|$ denotes the center-to-center unit vector between particles $1$ and $2$, and $\hat{\mathbf{r}}_i^\alpha$ is a unit vector from the center of particle $i$ to a patch $\alpha$ on its surface depending on the individual orientation $\Omega_i$. 
  Hence, the product of $\phi_\text{sw}$ and $G$ defines interactions between particles such that it is attractive only if two patches are within a small distance and are orientated to each other properly, depending on the patch opening angle $\theta_c$. A sketch visualizing the interaction of patchy particles is depicted in Fig. \ref{Fig:snapshotKF}. 
  Throughout this work, we assume $\delta = 0.119\sigma$, $\cos(\theta_c) = 0.895$, and the number of patches is $M = 4$. In order to compare with theoretical results, we performed Monte-Carlo simulations based on open-source code \cite{Rovigatti2018}, where the patches are distributed in a tetrahedral arrangement on the particles surfaces. Implementation details may be found in Appendix \ref{SubSec:MCSimulationDetails}.

  \subsection{Wertheim theory for patchy particles} \label{SubSec:TPT1}
  
  \begin{figure}[t] 
  	\centering
  	\includegraphics[width = 8cm]{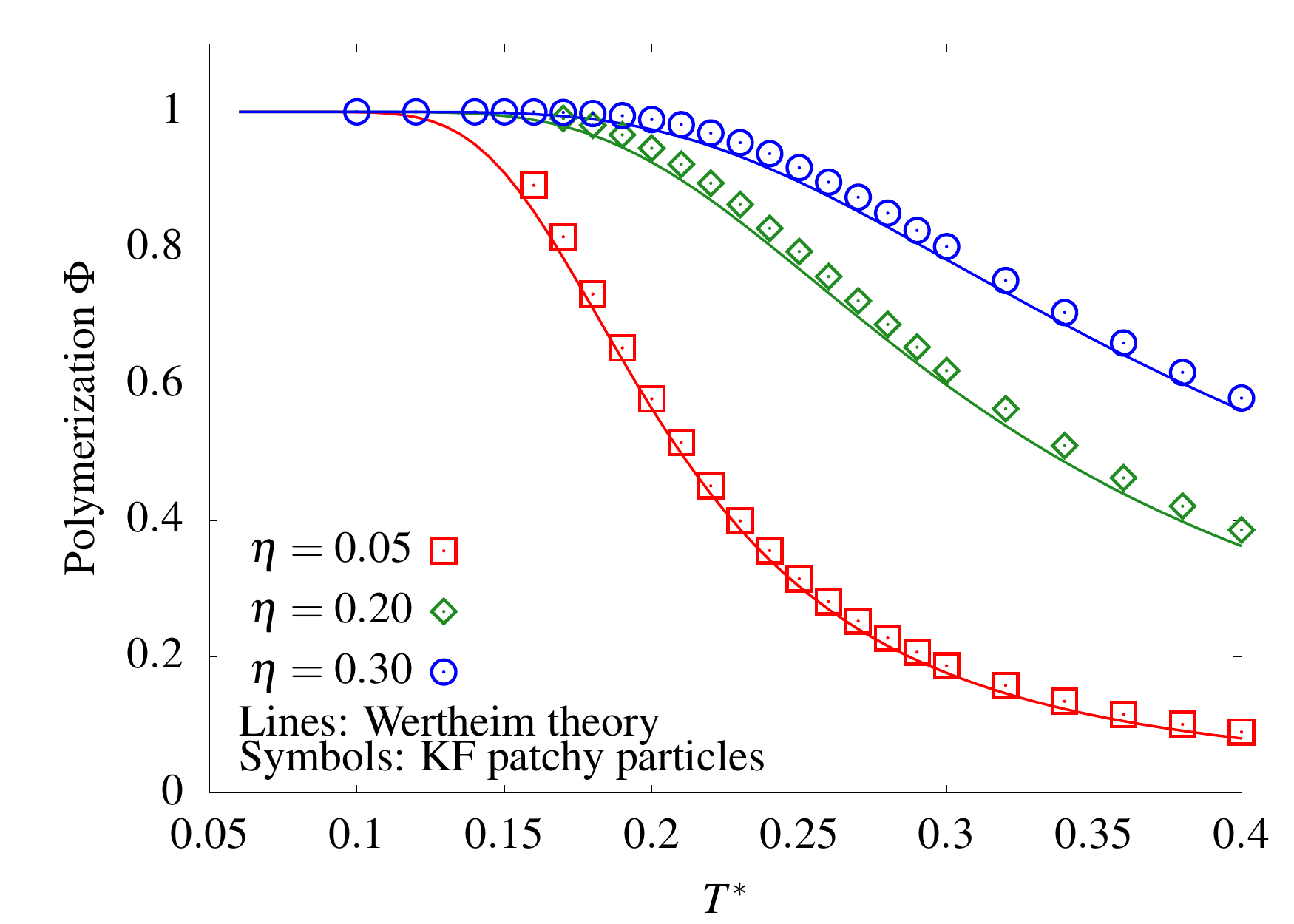}
  	\caption{Polymerization $\Phi$ as a function of $T^* = k_B T/\varepsilon$ for packing fractions $\eta = 0.05$ (red), 0.2 (green), and 0.3 (blue). Individual symbols show results from MC simulations and solid lines are predictions by TPT1. } \label{Fig:PolymerizationKF}
  \end{figure} 
  
  For patchy interactions, Wertheim has developed a successful thermodynamic perturbation theory \cite{WertheimI1984, WertheimII1984, WertheimIII1986, WertheimIV1986, Jackson1988}. In its first order formulation, the theory assumes that all $M$ patches on each particle are independent, where each patch can be engaged in precisely one bond connecting two particles. Furthermore, no information regarding the geometrical arrangements of the patches is included in the approach, and loop-forming structures such as rings are neglected. 
Note that the previously chosen values for $\delta$ and $\theta_c$ are in agreement with the assumption of maximum one
bond per patch.
  The contribution to the free energy density due to bond formation between particles can be written as \cite{Jackson1988}
  \begin{equation} \label{EqF_bond}
  \frac{\beta f_\text{bond}}{\rho_b} =  M \left[ \ln(1-p_b)  + \frac{p_b}{2} \right]\,,
  \end{equation}
  where $\rho_b = N/V$ denotes the number density, $\beta = 1/(k_B T)$ the inverse temperature, \textcolor{black}{$k_B$ is Boltzmann's constant}, and $p_b$ is the probability that an arbitrary site is bonded to another (fraction of bonded sites).
  For the present case of a one-component system with identical patches, $p_b$ is given by the following mass-action equation
  \begin{equation} \label{EqX}
  \frac{p_b}{(1-p_b)^2} = \rho_b M \Delta\,,
  \end{equation}
  where $\Delta$ is a measure for the interaction strength between two patches, and reads ($\langle \cdot \rangle_{\Omega_1, \Omega_2}$ denotes an angular average over all orientations of particles 1 and 2)
  \begin{align} \label{EqDelta}
  \Delta &=  \int \text{d}\mathbf{r}\, g_\text{hs}(r) \langle \exp(-\beta \phi_\text{patch}) - 1\rangle_{\Omega_1, \Omega_2} \notag \\
  &\approx v_b\, g_\text{hs}(\sigma^+)(e^{\beta\varepsilon} - 1)\,,
  \end{align}
  with the volume $v_b$ available for bonding given by
  \begin{equation} \label{Eq:BondingVolume}
  	v_b = \frac{4\pi\sigma^3}{3} \sin^4\left(\frac{\theta_c}{2}\right) \left[\left(1+\frac{\delta}{\sigma}\right)^3 -1\right]\,.
  \end{equation}
  The quantity $\Delta$ thus connects all geometrical and physical properties of the patches.
  The radial distribution function $g_\text{hs}(r)$ of the hard-sphere reference system has been approximated by its contact value $g_\text{hs}(\sigma^+)$, due to the short range of the site-site interaction. This contact value may be suitably approximated by the well-known Carnahan-Starling expression
  \begin{equation}
  g_\text{hs}(\sigma^+) = \frac{1-\frac{\eta}{2}}{(1-\eta)^3}\,,
  \end{equation}
\textcolor{black}{where $\eta = \pi \sigma^3 \rho_b / 6$ is the packing fraction of the fluid}.
  Essential bulk properties, such as the degree of polymerization $\Phi$ given by \cite{Flory1941, Bianchi2008}
  \begin{equation}
  	\Phi = 1 - (1 - p_b)^M\,,
  \end{equation} 
  are in good agreement with simulations. Theoretical results for this quantity as a function of $T^*$ are shown in Fig. \ref{Fig:PolymerizationKF} for packing fractions $\eta = 0.05$ (red), 0.2 (green), and 0.3 (blue). Individual symbols show simulation results, where the fractions of particles engaged in at least one bond have been accumulated for each state point.
  The gas-liquid binodal for $M = 4$ is, however, not in such good agreement with simulations. 
Here, Wertheim's theory particularly underestimates the densities of the coexisting liquid branch. 
The critical point for the present model parameters found in simulations is located at $(\eta_c,\, T^*_c) = (0.14,\, 0.168)$ \cite{Foffi2007}, and TPT1 in its present formulation predicts $(\eta_c,\, T^*_c)=(0.09,\, 0.161)$.

 \subsection{Density functionals for patchy particles} \label{SubSec:DFT}
 
 In classical density functional theory \cite{Evans1979}, the structure and thermodynamics of a fluid can be obtained from the grand-potential functional $\Omega[\rho]$ of the one-body density $\rho(\mathbf{r})$,
 \begin{equation} \label{Eq:grandPotentialFunctional}
 	\Omega[\rho] = F[\rho] + \int\text{d}\mathbf{r}\,\rho(\mathbf{r}) \left(V_\text{ext}(\mathbf{r}) - \mu\right)\,,
 \end{equation}
 which obeys a Euler-Lagrange equation in thermal equilibrium
 \begin{equation} \label{Eq:EulerLagrange}
 	0 = \left.\frac{\delta \Omega[\rho]}{\delta\rho(\mathbf{r})}\right|_{\rho=\rho_\text{eq}}\,,
 \end{equation} 
 \textcolor{black}{where $\rho_\text{eq}(\mathbf{r})$ means the density in equilibrium.}
 Here, $V_\text{ext}(\mathbf{r})$ denotes an arbitrary external potential acting on the fluid particles, and $\mu$ is the chemical potential (of the corresponding particle reservoir). Furthermore, $F[\rho] = F_\text{id}[\rho] + F_\text{ex}[\rho]$ is the intrinsic Helmholtz free-energy functional, which we can split into an exactly known ideal-gas part
 \begin{equation}
 	\beta F_\text{id}[\rho] = \int\text{d}\mathbf{r} \rho(\mathbf{r}) \left[\ln\left(\rho(\mathbf{r})\Lambda^3\right) - 1\right]\,,
 \end{equation}
 where $\Lambda$ is the thermal wavelength of the particles, and a so-called excess part $F_\text{ex}[\rho]$ which contains informations about the particle interactions. Of course, in general, this quantity 
is not known exactly as this would be equivalent to having access to the exact partition sum of the system. In limited cases, however, such as one-dimensional hard rods, exact expressions are available \cite{Percus1976, Vanderlick1989}. A key task in DFT is therefore to find reliable approximations for $F_\text{ex}[\rho]$ describing the underlying particle interactions as adequate as possible. Equation \eqref{Eq:EulerLagrange} yields an implicit equation for the density profile $\rho_\text{eq}(\mathbf{r})$ in equilibrium,
 \begin{equation} \label{Eq:densityProfileEq}
 	\rho_\text{eq}(\mathbf{r}) = \rho_b \exp\left(-\beta V_\text{ext}(\mathbf{r}) + c^{(1)}(\mathbf{r}) + \beta\mu_\text{ex}\right)\,,
 \end{equation}
 in which the chemical potential $\mu = \mu_\text{id} + \mu_\text{ex}$ was split into ideal-gas and excess contributions. The quantity $c^{(1)}(\mathbf{r})$ is the so-called one-body direct correlation function and is defined as 
 \begin{equation}
 	c^{(1)}(\mathbf{r}) := - \frac{\delta \beta F_\text{ex}[\rho]}{\delta\rho(\mathbf{r})}\,.
 \end{equation} 
 Similar, the $n$-body direct correlation functions are given by
 \begin{equation} \label{Eq:Defc^{(n)}}
 	c^{(n)}(\mathbf{r}_1,...,\mathbf{r}_n) := -\frac{\delta^n \beta F_\text{ex}[\rho]}{\delta \rho(\mathbf{r}_1)\cdots\delta\rho(\mathbf{r}_n)}\,.
 \end{equation}
 \textcolor{black}{
In particular, the bulk pair direct correlation function, 
\begin{equation} \label{Eq:Defc(r)}
	c(r) \equiv c^{(2)}(r = |\mathbf{r} - \mathbf{r}'|) = - \left.\frac{\delta^2 \beta F_\text{ex}[\rho]}{\delta\rho(\mathbf{r})\delta\rho(\mathbf{r}')}\right|_{\rho_b}\,,
\end{equation}
plays an important role in liquid-state theory. Via the Ornstein-Zernike (OZ) relation \cite{HansenMcDonald2013}
\begin{equation} \label{Eq:OZ} 
g(r) = c(r) + \rho_b \int\text{d}\mathbf{r}\, (g(r) - 1) c(|\mathbf{r} - \mathbf{r}'|)\,,
\end{equation}
$c(r)$ is linked to the radial distribution function $g(r)$ of the fluid, which characterizes the bulk structure of the system.}

Provided that an explicit expression for $F_\text{ex}[\rho]$ is known, Eq. \eqref{Eq:densityProfileEq} numerically can be solved using 
fixed point iterations to find the equilibrium density profiles in an arbitrary external potential. In this work, $F_\text{ex}[\rho]$ itself consists of a hard-sphere contribution $F_\text{hs}[\rho]$, and a contribution $F_\text{bond}[\rho]$ describing the patch-patch interactions. More precisely, we employ a functional on basis of the weighted density functional introduced by Tarazona \cite{Tarazona1990}, and the accurate fundamental measure theory by Rosenfeld \cite{Rosenfeld1989, Roth2010}. Both formalisms are recapitulated in the following sections.

 \subsubsection{Weighted-density functional} \label{SubSubSec:WDA}
The WDA functional that we employ in this work is based on Tarazona's WDA \cite{Tarazona1990} for hard spheres as extended by Kim \textit{et al.} \cite{Kim1995}. The excess free energy functional is written as
\begin{equation}
\beta F_\text{ex}[\rho] = \int \text{d}\mathbf{r}\,\rho(\mathbf{r})\beta\Psi(\bar{\rho}(\mathbf{r}))\,,
\end{equation}
where $\Psi(\bar{\rho}(\mathbf{r}))$ is a local excess free energy density per particle of the homogeneous system. $\bar{\rho}(\mathbf{r})$ is a weighted density defined via 
\begin{equation}
\bar{\rho}(\mathbf{r}) = \int\text{d}\mathbf{r}' \rho(\mathbf{r}') \omega(|\mathbf{r}-\mathbf{r}'|, \widetilde{\rho}(\mathbf{r}))\,.
\end{equation}
In bulk one has the constraint that $\bar{\rho}(\mathbf{r}) = \widetilde{\rho}(\mathbf{r}) =  \rho(\mathbf{r}) = \rho_b$ and thus $\omega(r,\rho)$ has to satisfy
\begin{equation}
\int\text{d}\mathbf{r}\, \omega(r, \rho) = 1\,.
\end{equation}
The idea of Tarazona was that the weight function $\omega(r, \rho)$ may be determined by requiring that the resulting bulk pair direct correlation function $c(r)$ should obey the analytically known first- and second-order terms of the (formally exact) virial expansion of $c(r)$. To this end, the weight function is expanded in powers of the density,
\begin{equation} \label{Eq:expansionWeightFunctions}
\omega(r,\rho) = \omega_0(r) + \rho\omega_1(r) + \rho^2\omega_2(r)\,+\,...\,,
\end{equation}
giving rise to the equation
\begin{equation} \label{Eq:Quadratic_rhobar}
\bar{\rho}(\mathbf{r}) = \rho_0(\mathbf{r}) + \widetilde{\rho}(\mathbf{r})\rho_1(\mathbf{r}) + [\widetilde{\rho}(\mathbf{r})]^2\rho_2(\mathbf{r})\,+\,...\,,
\end{equation}
in which
\begin{equation}
\rho_i(\mathbf{r}) \equiv \int\text{d}\mathbf{r}'\,\rho(\mathbf{r}')\omega_i(|\mathbf{r}-\mathbf{r}'|)\,,\text{with}\,i\,=\,0\,,1\,,2\,,...
\end{equation}
Note that in Tarazona's original work $\widetilde{\rho}(\mathbf{r}) = \bar{\rho}(\mathbf{r})$, thus Eq. \eqref{Eq:Quadratic_rhobar} yields a quadratic equation for $\bar{\rho}(\mathbf{r})$ when truncated at second order in the density.
Kim \textit{et al.} alternatively suggested that the weighted densities $\widetilde{\rho}(\mathbf{r})$ may be obtained by the global average
\begin{equation}
\widetilde{\rho}(\mathbf{r}) = \int\text{d}\mathbf{r}' \rho(\mathbf{r}')\omega(|\mathbf{r}-\mathbf{r}'|, \rho_b)\,,
\end{equation}
which guarantees that the resulting bulk pair direct correlation function $c(r)$ still is given as in Tarazona's original WDA, 
\begin{align} \label{Eq:PairDirectCorrelationWDA}
c(r) = &-2\beta\Psi'(\rho_b)\omega(r, \rho_b) - \rho_b\beta\Psi''(\rho_b)\, (\omega \ast \omega)(r, \rho_b)\notag \\ 
&- 2\beta\rho_b\Psi'(\rho_b)\,(\omega\ast\omega')(r, \rho_b)\,,
\end{align}
where primes in Eq. \eqref{Eq:PairDirectCorrelationWDA} mean differentiation with respect to the density and $\ast$ denotes a three-dimensional convolution of the weight functions.
The form of Kim \textit{et al.} produces nearly indistinguishable results compared to the original Tarazona-WDA, but is computationally less intensive as it only requires the calculation of additional convolutions. In particular, one can employ the same weight functions $\omega_i(r)$ as in Ref. \cite{Tarazona1990}. The Fourier representation of the $\omega_i$ are given in Appendix \ref{SubSec:WDAweightFunctionsFourier}.
 
For the free energy density per particle of pure hard spheres the very accurate Carnahan-Starling expression is employed 
\begin{equation}
\beta\Psi_\text{cs}(\bar{\rho}(\mathbf{r})) = \frac{4\bar{\eta}(\mathbf{r})-3\bar{\eta}(\mathbf{r})^2}{(1-\bar{\eta}(\mathbf{r}))^2}\,,
\end{equation}
where $\bar{\eta}(\mathbf{r}) = \pi \sigma^3 \bar{\rho}(\mathbf{r})/6$.
Although the above approach has been derived for the pure hard-sphere fluid, Segura \textit{et al.}  were the first who used this WDA formalism to obtain a density functional for associating fluids, where \cite{Segura1997, GnandelasHeras2012}
\begin{align}
\beta\Psi(\bar{\rho}(\mathbf{r})) &= \beta\Psi_\text{cs}(\bar{\rho}(\mathbf{r}))\notag \\ &+ M\left[\log(1 - p_b(\bar{\rho}(\mathbf{r}))) + \frac{p_b(\bar{\rho}(\mathbf{r}))}{2}\right]\,.
\end{align} 
As can be seen, the bonding contribution in the inhomogeneous case is assumed to be of Wertheim's 
form, Eq.~(\ref{EqF_bond}), where the bulk density dependence of the bonding probability has been replaced by a dependence
on the weighted density $\bar{\rho}(\mathbf{r})$.

 \subsubsection{Fundamental measure functional} \label{SubSubSec:FMT}
 While Tarazona's WDA for hard spheres has proven to yield reliable predictions even at very high densities, the most recent form of density functionals for hard bodies are based on Rosenfeld's FMT \cite{Rosenfeld1989, Roth2010}. Besides the 
gratifying observation that it performs even better at very high fluid packing fractions compared to simulations, a cornerstone is posed by the fact that it generates the Percus-Yevick (PY) bulk pair correlation function $c_\text{py}(r)$ as an output (without using it as an input).
 At the center of FMT is the observation that the Mayer-$f$ function of hard spheres, $f(r) = -\Theta(\sigma - r)$, \textcolor{black}{where $\Theta(\cdot)$ is the Heaviside step function}, can be decomposed into convolutions of weight functions $\omega_\alpha(\mathbf{r})$ that characterize the geometry of the particle. As a result, the 
exact virial expansion up to second order of the free-energy functional can be expressed in terms of weighted densities $n_\alpha(\mathbf{r})$,
 \begin{align}
 \beta F_\text{ex}[\rho] &= -\frac{1}{2}\int\text{d}\mathbf{r}\text{d}\mathbf{r}'\,\rho(\mathbf{r})\rho(\mathbf{r}')f(|\mathbf{r}-\mathbf{r}'|)\notag \\	
 &= \int\text{d}\mathbf{r} \{n_0(\mathbf{r})n_3(\mathbf{r}) + n_1(\mathbf{r})n_2(\mathbf{r}) - \mathbf{n}_1(\mathbf{r})\cdot \mathbf{n}_2(\mathbf{r})\}\,,
 \end{align}
 where
 \begin{equation}
 n_\alpha(\mathbf{r}) = \int\text{d}\mathbf{r}'\,\rho(\mathbf{r}')\omega_\alpha(\mathbf{r}-\mathbf{r}')\,.
 \end{equation}
 For instance, $n_3(\mathbf{r})$ is a local packing fraction given by
 \begin{equation}
 n_3(\mathbf{r}) = \int\text{d}\mathbf{r}'\,\rho(\mathbf{r}')\Theta(\sigma/2 - |\mathbf{r}-\mathbf{r}'|)\,,
 \end{equation}
 which reduces to the global packing fraction $\eta$ in the uniform fluid.
 For the extrapolation to higher densities Rosenfeld employed the ansatz
 \begin{equation}
 \beta F_\text{ex}[\rho] = \int\text{d}\mathbf{r}\,\Phi_\text{hs}(\{n_\alpha(\mathbf{r})\})\,,
 \end{equation}
 inspired by the exact result for hard rods in one dimension \cite{Percus1976}. Along with dimensional analysis, and a connection to thermodynamics relating the chemical potential when inserting an infinite large sphere into the fluid to the bulk pressure $\beta p$, Rosenfeld was able to derive a closed solution for $\Phi_\text{hs}(\{n_\alpha\})$ which reads
 \begin{align}
 \Phi_\text{hs}(\{n_\alpha\}) = &-n_0\ln(1-n_3) + \frac{n_1 n_2 - \mathbf{n}_1\cdot\mathbf{n}_2}{1-n_3} \notag \\ &+ \frac{n_2^3 - 3 n_2 \mathbf{n}_2\cdot\mathbf{n}_2}{24\pi(1-n_3)^2}\,.
 \end{align}
 This functional subsequently has been further improved by Tarazona to describe hard-sphere crystals \cite{Tarazona2000}, and later by Roth \textit{et al.} \cite{Roth2002} and Hansen-Goos \cite{HansenGoos2006WB2} making use of more accurate equations of state. \textcolor{black}{Note that any density functional which makes use of FMT-type weighted densities yields the following generic form for the bulk pair direct correlation function
 	\begin{equation}
 	c(r) = - \sum_{\alpha,\beta} \left.\frac{\partial^2 \Phi}{\partial n_\alpha \partial n_\beta}\right|_{\rho_b} (\pm \omega_\alpha \ast \omega_\beta)(r)\,,
 	\end{equation}
 where the negative sign holds for the vector-type weighted densities.
 }
 
In their work, Yu and Wu extended the bonding contribution to the inhomogeneous fluid via a weighted-density approximation of Eqs. \eqref{EqF_bond}, \eqref{EqX}, and \eqref{EqDelta} incorporating the weighted densities $n_\alpha(\mathbf{r})$ of FMT. More specifically, the excess free-energy functional has the form \cite{YuWu2002}
\begin{equation}
	\beta F_\text{ex}[\rho] = \int\text{d}\mathbf{r}\, \left[\Phi_\text{hs}(\{n_\alpha(\mathbf{r})\}) + \Phi_\text{bond}(\{n_\alpha(\mathbf{r})\})\right]\,,
\end{equation}
where
\begin{equation}
	\Phi_\text{bond}(\{n_\alpha\}) = M\, n_0(\mathbf{r})\, \xi(\mathbf{r})^q \left[\ln\left(1-p_b(\mathbf{r})\right) + \frac{p_b(\mathbf{r})}{2}\right]\,,
\end{equation}
with $p_b(\mathbf{r})$ being now a position-dependent function via the weighted densities $n_\alpha(\mathbf{r})$ given by
\begin{equation}
	\frac{p_b(\mathbf{r})}{(1-p_b(\mathbf{r}))^2} = n_0(\mathbf{r})\,\xi(\mathbf{r})^q\, M\,\Delta(\{n_\alpha(\mathbf{r})\})\,.
\end{equation}
The quantity $\Delta = v_b (e^{\beta\varepsilon} - 1) g_\text{hs}(\{n_\alpha(\mathbf{r})\})$ depends on weighted densities through the generalized contact value of hard spheres, given by
\begin{align}
g_\text{hs}(\{n_\alpha\}) = \frac{1}{1-n_3} + \frac{\sigma n_2 \xi^q}{4(1-n_3)^2} 
+ \frac{\sigma^2 n_2^2 \xi^q}{72(1-n_3)^3}\,.		
\end{align}
Importantly, the factor $\xi = 1 - \mathbf{n}_2\cdot\mathbf{n}_2/n_2^2$ has been incorporated purely phenomenologically by Yu and Wu, probably to obtain better results in comparison to computer simulations. Note that in this work we allow for a slightly more general form for $\xi$ in powers of $q$. This, of course, does not affect the bulk form of Wertheim's theory, since in uniform systems $\xi^q = 1$ for all values of $q$ due to the fact that the volume integral over the vector-type weight functions vanishes. Yu and Wu assumed $q = 1$; however, as we will see later, results in the present work significantly can be improved when employing $q = 3$. To the latter we will refer to as the mFMT (modified FMT) functional, whereas the original approach with $q = 1$ is referred to as the FMT functional. 
Implementation details for the DFT may be found in Appendix \ref{Sec:DFTImplementationDetails} including Fourier representations of the weight functions employed for the WDA and for the FMT.

\subsection{Test particle approach} \label{SubSec:TestParticleApproach}

\begin{figure}[t] 
	\centering
	\includegraphics[width = 8cm]{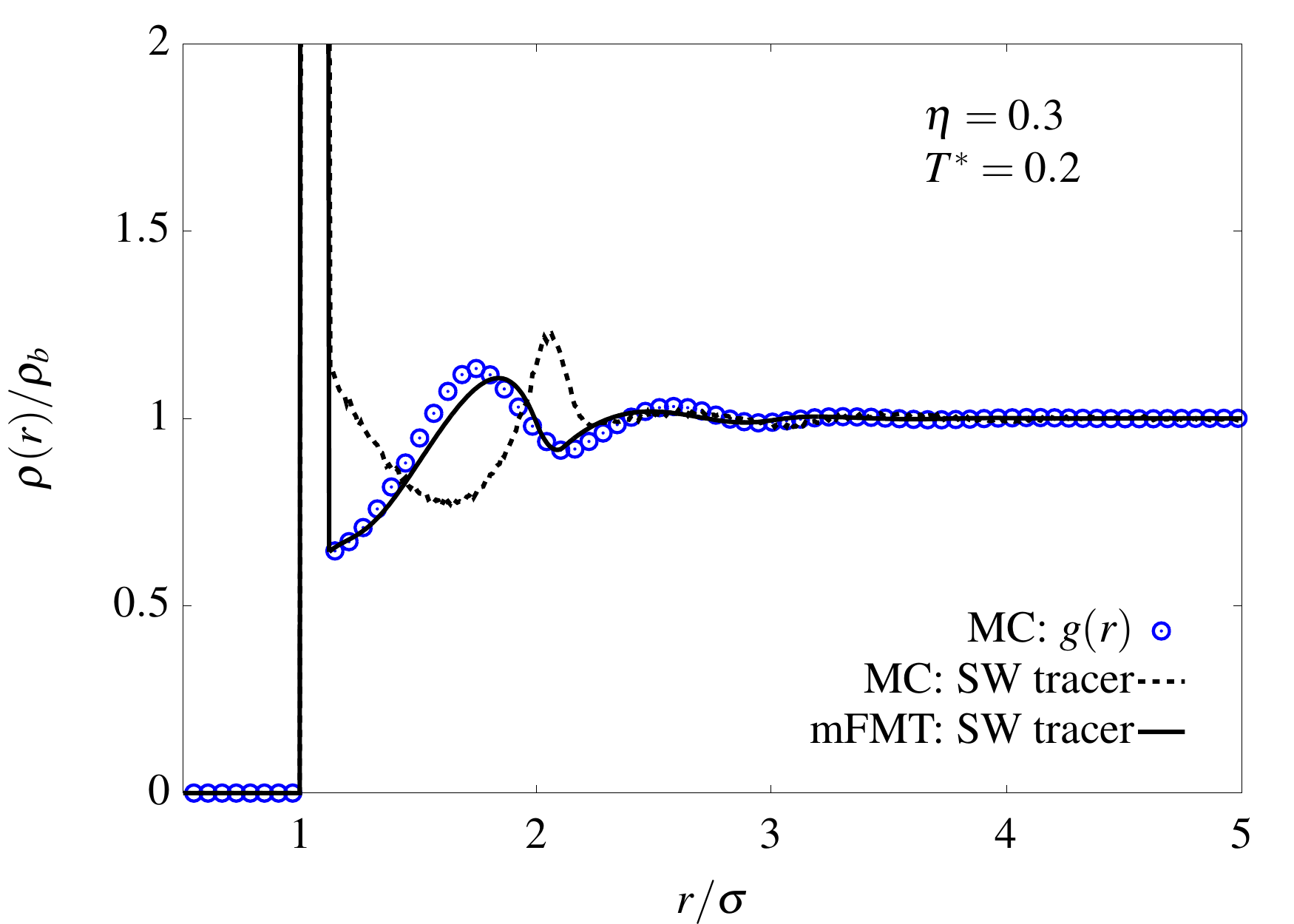}
	\caption{Radial distribution function $g(r)$ obtained from MC simulations of patchy particles (individual symbols) at packing fraction $\eta = 0.3$ and temperature $T^* = 0.2$. The dashed line shows the corresponding MC result for the density around a square-well type test particle. The solid line plots a DFT result around an external potential according to 
Eq. \eqref{Eq:TestParticlePotential} with $\varepsilon_\text{sw}$ chosen to match the first peak in the simulated
$g(r)$ (not visible on the scale of the graph).} \label{Fig:effective_testparticle}
\end{figure}   

Within DFT, the density distribution around a test particle can readily be calculated by applying a suitable external potential $V_\text{ext}(\mathbf{r})$. In particular, if $V_\text{ext}(\mathbf{r})$ is equivalent to the interaction potential of the particles, the resulting density profile divided by the bulk density is equivalent to the bulk radial distribution function $g(r)$ of the fluid; this is the well-known Percus' test particle approach \cite{HansenMcDonald2013}. 
In the present context of particles with directional interactions this, however, poses a challenging task. Any density functional which employs the bulk formulation of the Wertheim theory and makes use of purely spherical weight functions cannot depend on the particle orientations. This makes it very hard to realize an \textit{exact} realization of Percus' test particle theory as this would require a proper coupling of the orientational character of the pair potential and an orientationally independent \textcolor{black}{ensemble-averaged} density profile of the particles. 

In this work, we propose a method to approximatively calculate radial distribution functions from DFT as follows. We calculate the density profiles around a particle that interacts with the fluid via a short-ranged spherical square-well (SW) potential, where the range is set equal to the patch-patch interaction range $\delta$. More precisely, the external potential acting on the fluid is given by
\begin{align} \label{Eq:TestParticlePotential}
V_\text{ext}(r) 
&= \begin{cases}
\infty~~~~~;~~~ &r < \sigma \\
-\varepsilon_\text{sw} ~~;~ \sigma \leq &r < \sigma+\delta \\
~~0 ~~~~~;~&r \geq \sigma+\delta\,,
\end{cases}
\end{align}
where $\varepsilon_\text{sw}$ sets the interaction energy between test particle and the surrounding particles. 

In simulations, such a spherical tracer \textcolor{black}{(in view of Percus' test particle approach this means that the tracer does not know about the orientational character of the interaction between the surrounding particles, and interacts with the latter via a spherically-symmetric square-well pair potential)} will influence the local ordering of the particles within the first few coordination shells and will not result in the radial distribution function $g(r)$. Indeed, in order to minimize configurational energy, we expect that the particles surrounding a SW test particle prefer such orientations in which their patches are not face-to-face with the test particles as this would result in a decrease of bonding possibilities. This is demonstrated in Fig. \ref{Fig:effective_testparticle}, in which a radial distribution function $g(r)$ obtained from MC simulations of patchy particles (blue symbols) at packing fraction $\eta = 0.3$ and temperature $T^* = 0.2$ is compared to the density profile around a spherical attractive test particle (black dashed line). The potential depth $\varepsilon_\text{sw}$ is chosen such that the height of the first peak matches the peak height found in the $g(r)$. The second peak in $g(r)$ at $r \approx 1.75 \sigma$, which is typical for directional fluids exhibiting tetrahedral networks, is not found in the correlations around the tracer, but occurs at $r \approx 2\sigma$ indicating a local distortion of the bulk structure. \\

However, such a distortion of the local structure mainly driven by distinguished particle orientations cannot be captured by the present density functionals, since, as mentioned in the first paragraph of this subsection, they do not consider preferred orientations explicitly. In particular, this may be interpreted as if the theory assumes that (bulk) particle orientations are unaffected by a given external potential. Thus, for the present situation describing the bulk structure of a directional fluid, one may benefit from the orientational independence of the theory and for suitably chosen values of the potential depth $\varepsilon_\text{sw}$, Eq. \eqref{Eq:TestParticlePotential} may 
mimic an \textit{effective} (orientationally averaged) patch-patch interaction potential.
Note that if $\varepsilon_\text{sw}$ is chosen such to fit the first peak of $g(r)$, the first shell around the
tracer particle contains as many particles as a true test particle, and according to the argument given before,
in DFT the first shell corresponds to particles properly oriented with their patches towards the tracer particle. \textcolor{black}{This situation can also be modeled explicitly in simulations: If the spherical interaction introduced by the tracer acts only on the patches of the surrounding particles,  then $\varepsilon_\text{sw}$ can be chosen such that the resulting density profiles deviate only marginally from the real $g(r)$.}

 For illustration, when calculating the density profiles around an external potential according to Eq. \eqref{Eq:TestParticlePotential} making use of the mFMT functional, we find excellent agreement between the mFMT functional (black solid line in Fig. \ref{Fig:effective_testparticle}) and the radial distribution function $g(r)$ from simulations. 
Nevertheless, it is important to note that this approach should be most reliable in case of isotropically distributed patches, which is in best accordance with the assumption of orientationally independent density functionals. For instance, we have verified that the $g(r)$ a fluid of chains (with two sites per particle, i.e. $M=2$) is not adequately described with the present approach: The correlation peaks which in this case occur roughly at integer values of $\sigma$ in simulations, are still located at non-integer values in the DFT results.

  \begin{figure}[t] 
  	\centering
  	\includegraphics[width = 8cm]{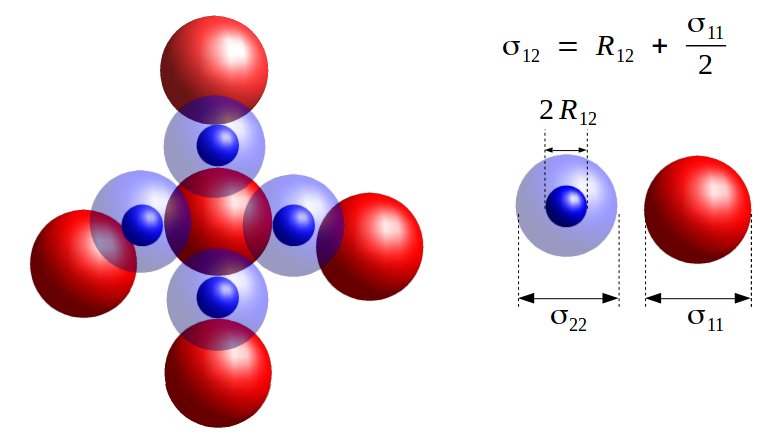}
  	\caption{Realization of a finite valence in the floating--bond model:
     The hard interaction radius between particles and bonds $\sigma_{12}$ is independent from
    $\sigma_{11}$ and $\sigma_{22}$. The large $\sigma_{22}$ (transparent blue spheres)
    prevents that more than 4 bonds can bind to a particle (red sphere).} \label{Fig:floatingbonddef}
  \end{figure} 

    \begin{figure*}[t] 
    	\centering
    	\includegraphics[width = 8cm]{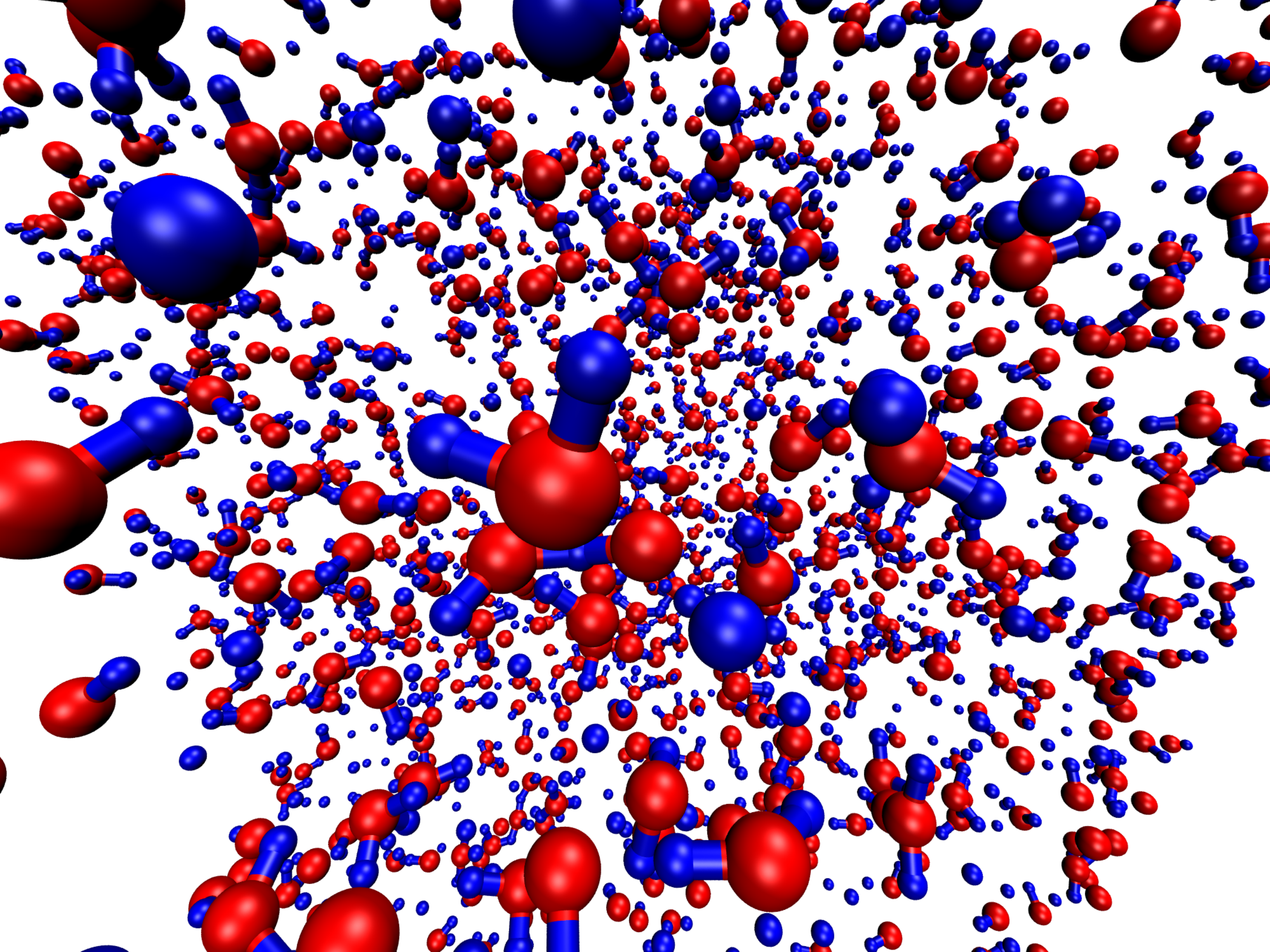}
    	\includegraphics[width = 8cm]{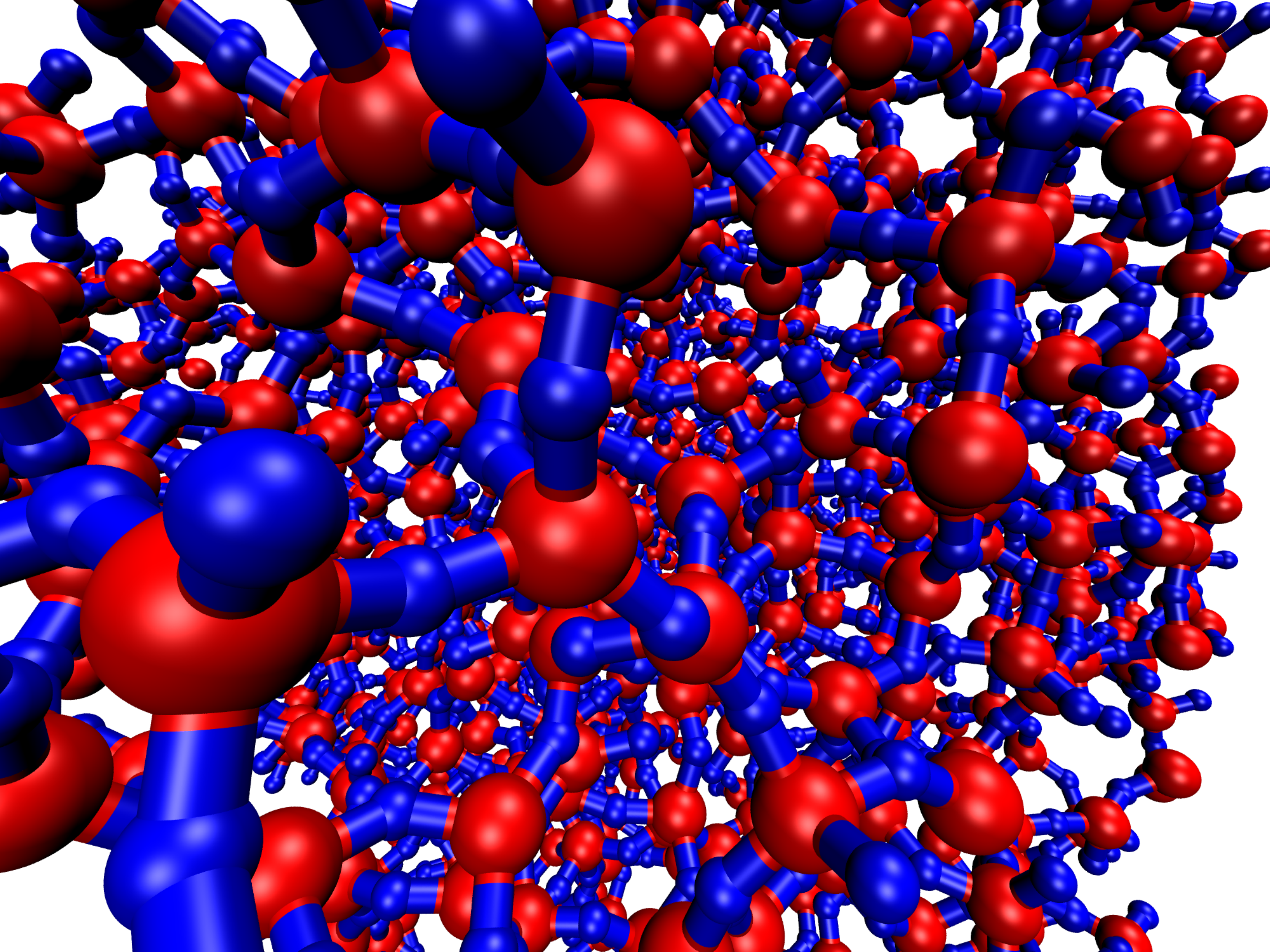}
    	\caption{Simulation snapshots \cite{HUMP96} of the floating-bond model at packing fraction $\eta = 0.05$ and temperature $T^*_{12} = 0.2$ (left) and $\eta = 0.3$ and $T^*_{12} = 0.1$ (right). The corresponding degree of polymerization is $\approx 20$\% and $\approx 99$\%, respectively. Blue spheres are the bonds and red spheres show the particles.} \label{Fig:snapshot_FB}
    \end{figure*} 

\subsection{Floating-bond model} \label{SubSec:FloatingBonds}
 
  Alternatively to patchy particles, in the floating-bond model directional interactions are generated via a non-additive binary mixture of small particles with diameter $\sigma_{22}$ (floating bonds, or bonds) and larger particles (or just \lq particles\rq\,) with diameter $\sigma_{11}$ interacting via purely isotropic potentials \cite{Zaccarelli2007}, 
see Fig.~\ref{Fig:floatingbonddef}. 
While floating bonds and large particles themselves repel each other, bonds and particles feel an additional attraction. The total interaction potential of the system reads \cite{Zaccarelli2007}
  \begin{align} \label{Eq:FloatingBondPotentialDiscontinuous}
  \phi_{11}(r) &= 
  \begin{cases}
  \infty~~;~~r < \sigma_{11} \\
  0~~~~;~r > \sigma_{11} \notag\,,\\
  \end{cases}
  \\
  \phi_{22}(r) &= 
  \begin{cases}
  \infty~~;~~r < \sigma_{22} \\
  0~~~~;~r > \sigma_{22}\,, \\
  \end{cases}
  \\
  \phi_{12}(r) &=
  \begin{cases}
  \infty~~~~~;~~r < \sigma_{12} \\
  -\varepsilon_{12} ~~;~~\sigma_{12} < r < \sigma_{12} + \delta_{12} \notag \\
  0~~~~~~~;~~r > \sigma_{12} + \delta_{12}\,.
  \end{cases}
  \end{align}
  In order to obtain a maximum number of floating bonds per particle equal to four and maximal two particles per bond (to model the situation of a tetravalent fluid) one may use $\sigma_{22} = 0.8\sigma_{11}$, $\sigma_{12} = 0.55\sigma_{11}$, and $\delta_{12} = 0.03\sigma_{12}$ \cite{Zaccarelli2007}. The effective range in which two particles are linked via a bond is thus $\sigma_{11} < r < 2(\sigma_{12} + \delta_{12}) = 1.16\sigma_{11}$. The number of bonds is chosen such that in the ground state all particles are connected to four other particles, which can be achieved by choosing a number of bonds twice as much as particles being present in the system.   
In this work we do not precisely employ the discontinuous potentials of Eq. \eqref{Eq:FloatingBondPotentialDiscontinuous} as we use Brownian dynamic methods to simulate the floating-bond model. Instead, we employ a slightly adapted version with continuous potentials as introduced by some of the authors in Ref. \cite{Bleibel2018}. Details can be found in the Appendix \ref{SubSec:FloatingBondSimulationDetails}. 
For the floating-bond model, all packing fractions refer only to the large particles. 
Snapshots of typical configurations are shown in Fig. \ref{Fig:snapshot_FB} for a rather low packing fraction $\eta = 0.05$ and high temperature $T^*_{12} = k_B T/\varepsilon_{12}$ (left), and a state point close to the optimal network region \cite{Zaccarelli2007} at $\eta=0.3$ and $T^*_{12} = 0.1$ (right). \textcolor{black}{To locate the critical point, we computed pressure isotherms
in the canonical ensemble and estimated the position of the spinodal line. From that we found that the critical point is approximately located at ($\eta_c$, $T_{12,c}^*$) = (0.1, 0.105) which
is at a slightly higher temperature than the critical point
estimated by Zaccarelli et al. \cite{Zaccarelli2007} (using the same method) who found $T_{12, c}^*$= 0.095  using the hard potential version. No finite size scaling has been performed.}

  Interestingly, the approximations introduced by TPT1 may be viewed as conceptual similarities shared with the floating-bond model: Besides the single-bond condition, and the maximum number of bonded neighbors, in both models there is no rigid geometrical constraint regarding the position of sites. Moreover, Wertheim's first-order theory is not sensitive for cases in which bonding at one site would block bonding at another. In the floating-bond model, such situations are also minimized due to the fact that a bond is flexible on the particles surface: Although if a existing link between two large particles (mediated by some bond connecting them) would in principle impede another connection due to steric incompatibilities, the other bond mediating the other connection may simply rearrange on the particles surface such that a connection becomes possible. 
   \begin{figure}[t] 
   	\centering
   	\includegraphics[width = 8cm]{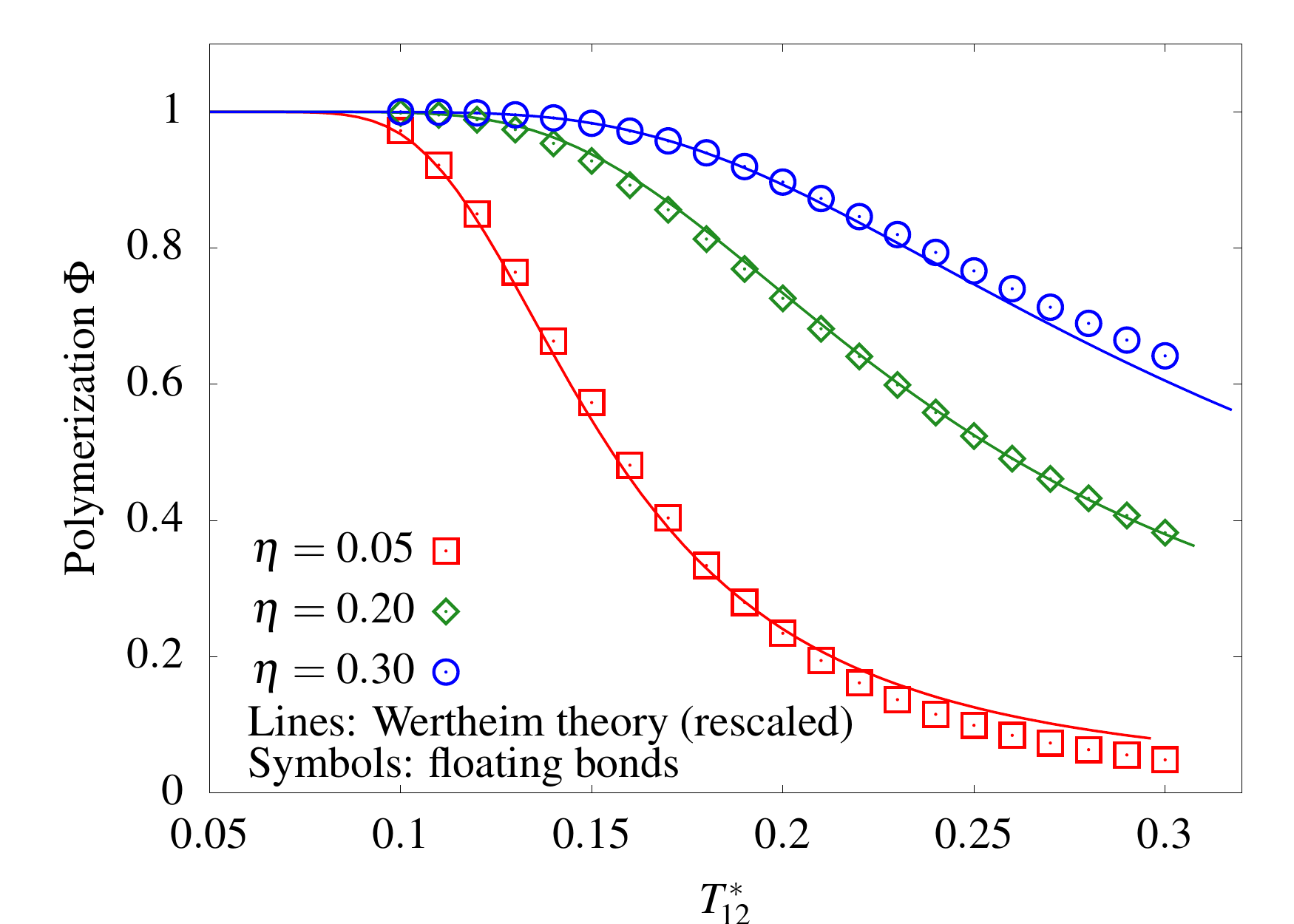}
   	\caption{Polymerization $\Phi$ in the floating-bond model as a function of the reduced temperature $T^*_{12} = k_B T/\varepsilon_{12}$ for packing fractions $\eta = 0.05$ (red), 0.2 (green), and 0.3 (blue). Individual symbols show results from BD simulations, and solid lines are predictions by TPT1 with a temperature rescaling (see text). } \label{Fig:PolymerizationFB}
   \end{figure}       
  However, it is also clear that the overall situation generated by the floating-bond model cannot directly be compared to systems of pure patchy particles. For instance, the floating bonds introduce additional entropic contributions into the system. This naturally yields suspicions when directly (i.e., parameter-free) comparing e.g. gas-liquid phase diagrams.
However, $\Delta$ is the only parameter in TPT1 for the bonding contribution to the free energy, see Eq.~(\ref{EqDelta}).
Here, we fit $\Delta$ to the degree of polymerization $\Phi$. To facilitate a comparision between the floating-bond model
and the Kern--Frenkel model, we express the change in $\Delta$ by simply a temperature rescaling while keeping
the KF value for the bonding volume $v_b$. This procedure works quite well as can be seen in 
Fig. \ref{Fig:PolymerizationFB}.
Individual symbols in Fig. \ref{Fig:PolymerizationFB} show $\Phi$ as a function of the reduced temperature $T^*_{12}$ as obtained from simulations of floating bonds for the same volume fractions as in Fig. \ref{Fig:PolymerizationKF}. The solid lines display the corresponding results from TPT1 with a suitable temperature rescaling. The best agreement was obtained with a scaling of $T^*_{12} = T^*/1.3$ for all three packing fractions, i.e. the TPT1 effective temperature in the floating-bond 
model is about 30\% larger.

  \begin{figure}[t] 
  	\centering
  	\includegraphics[width = 8cm]{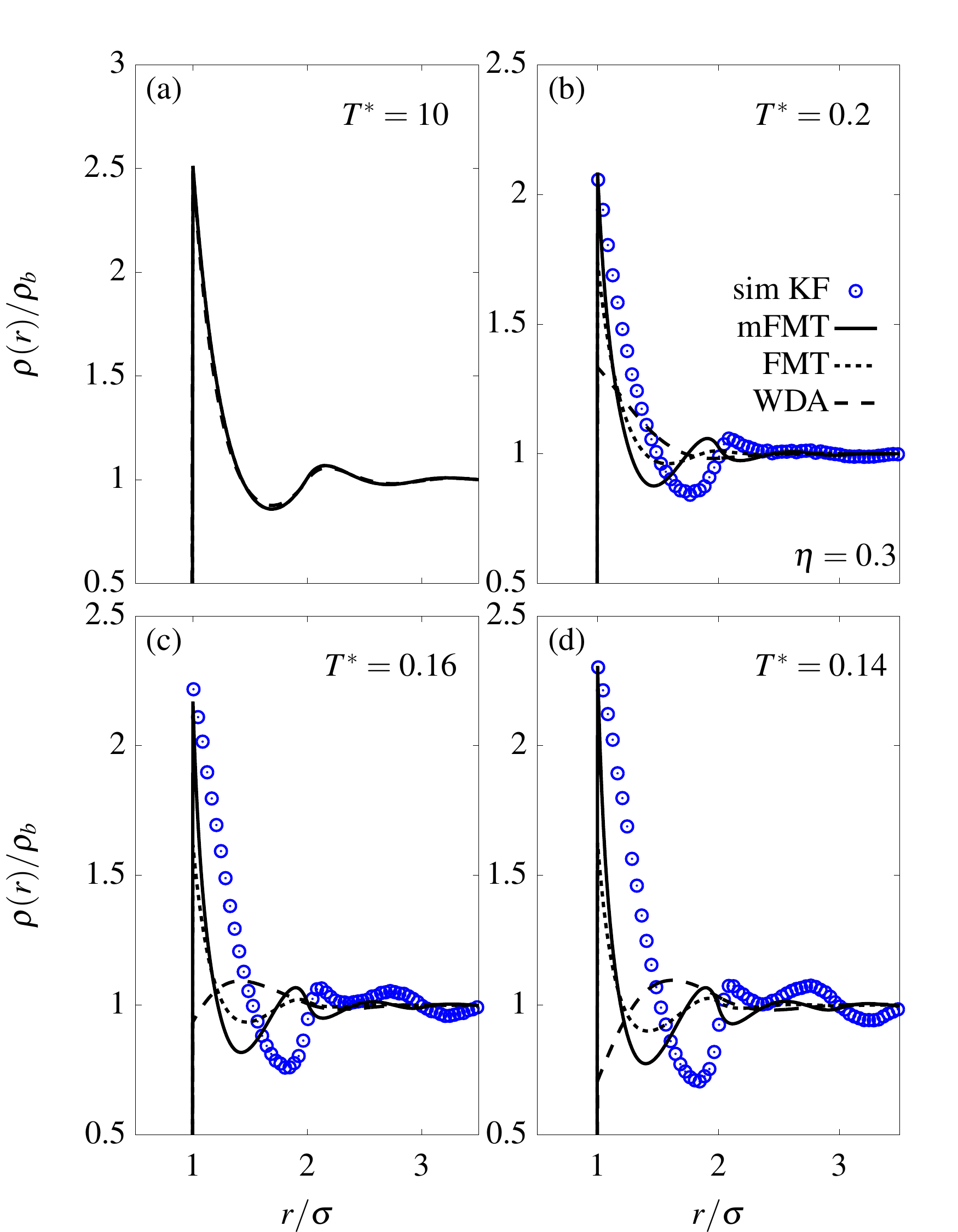}
  	\caption{Normalized density profile $\rho(r)/\rho_b$ around a hard spherical tracer particle ($\varepsilon_\text{sw} = 0$) with same hard-core diameter $\sigma$ as the fluid particles. The packing fraction is $\eta = 0.3$. The temperatures $T^*$ are 10.0 (a), 0.2 (b), 0.16 (c), and 0.14 (d). The solid lines are the results as borne out by the mFMT-functional, short-dashed by the FMT, and long-dashed lines by the WDA. The individual symbols in (b)-(d) show results from MC simulations.} \label{Fig:WDA_vs_FMT_hard_solute}
  \end{figure}  

\section{Results} \label{Sec:Results}
 
\subsection{Density around hard spherical tracer} \label{SubSec:CompWDAFMT}

We start with comparing the behavior of the WDA, FMT, and the mFMT functional to simulation results for the structure around a hard spherical tracer of diameter $\sigma$, i.e., $\varepsilon_\text{sw} = 0$ in Eq. \eqref{Eq:TestParticlePotential}. The results can be found in Fig. \ref{Fig:WDA_vs_FMT_hard_solute}, for temperatures $T^*$ = 10.0 (a), 0.2 (b), 0.16 (c), and 0.14 (d). The packing fraction is $\eta = 0.3$, which yields a degree of polymerization $\Phi$ ranging from nearly zero at $T^* = 10.0$ where particles behave as hard spheres, to nearly fully bonded states with $\Phi \approx 1$ at low temperatures, cf. also Fig. \ref{Fig:PolymerizationKF}. 
Lines in Fig. \ref{Fig:WDA_vs_FMT_hard_solute} correspond to results from mFMT (solid), FMT (short-dashed), and WDA (long-dashed), whereas individual symbols show simulation results. We have verified that all DFT implementations fulfilled the Gibbs adsorption theorem providing a proof of internal consistency \cite{Roth2010}. 

While in the hard-sphere limit all three curves are nearly indistinguishable, qualitative differences between (i) the functionals and (ii) with the simulations emerge when association between the particles increases. The WDA predicts a continuous decrease of the contact density $\rho(\sigma^+)$ while simultaneously a broad correlation peak at $r \lesssim 1.6\sigma$ emerges, whose intensity does not further increase for $T^* < 0.16$. This behavior bears resemblance to the results reported in Ref. \cite{GnandelasHeras2012}, where density profiles at a planar hard wall have been considered for trivalent patchy particles. 

  \begin{figure*}[t] 
  	\centering
  	\includegraphics[width = 18cm]{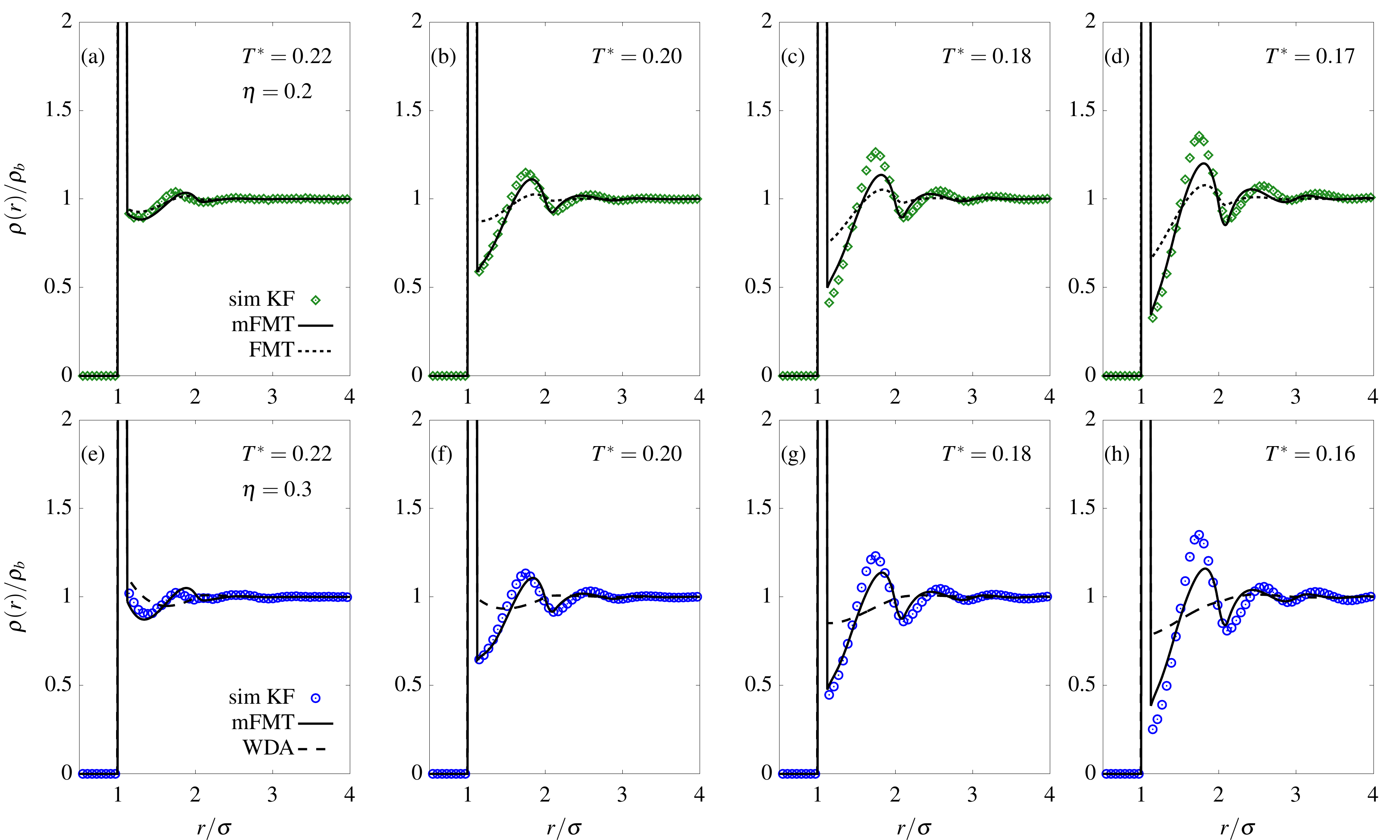}
  	\caption{Density profiles from mFMT (black solid lines) fitted via the effective test-particle route to simulations (individual symbols) of KF patchy particles for $\eta = 0.2$ (upper panel) and 0.3 (lower panel) at temperatures $T^* = 0.22$ [(a), (e)], 0.2 [(b), (f)], 0.18 [(c), (g)], 0.17 [(d)], and 0.16 [(h)]. The short-dashed lines in the upper panel correspond to the original FMT, while the long-dashed lines in the bottom panel show results from the 
  		WDA functional.} \label{Fig:FMT_vs_Patchy_eta02}
  \end{figure*} 

In contrast, the contact values given by the mFMT and the FMT functional, respectively, increase slightly upon cooling. Importantly, only the mFMT approach provides an accurate representation of the contact density and the structural intensity in comparison to simulation results. However, the density peaks are out of phase with the latter. Facing the discussion in the previous Sec. \ref{SubSec:TestParticleApproach}, this may also be traced back to the fact that the theory cannot account for orientational inhomogeneities as necessarily are introduced by a spherical hard test particle. 
The particles directly surrounding the tracer will prefer to orient themselves such that the number of possible bonds is maximized, i.e. it is unlikely that a patch is oriented face-to-face with the test particle; this gives rise to that the second peak in $\rho(r)$ is located at $r \approx 2\sigma$ instead of $r \approx 1.75\sigma$ as would be the case within a perfect bulk situation. The third peak located at $r \approx 2.7\sigma$ [most prominent visible for the lowest $T$ considered here, cf. Figs. \ref{Fig:WDA_vs_FMT_hard_solute} (c) and (d)] is then associated with the structural information of a tetrahedral network starting in the third coordination shell around the hard tracer. The (m)FMT predicts that the bulk order is not significantly distorted by the presence of the hard tracer particle, and thus the second peak arises at $r < 2\sigma$.

 \subsection{Effective radial distribution functions} \label{SubSec:CompSimDFT}

In this section, we examine whether the test-particle approach described in Sec. \ref{SubSec:TestParticleApproach} can be applied to \textit{effectively} describe radial distribution functions $g(r)$ and compare to simulation results. We start with comparison between DFT and MC simulations of Kern-Frenkel-type patchy particles. In Fig. \ref{Fig:FMT_vs_Patchy_eta02} we plot radial distribution functions obtained from simulations (individual symbols) compared to DFT results employing the effective test-particle method for $\eta = 0.2$ (upper panel) and $0.3$ (lower panel) at temperatures $T^* = 0.22$ [(a) and (e)], 0.2 [(b) and (f)], 0.18 [(c) and (g)], 0.17 [(d)] and 0.16 [(h)] (remember, the critical temperature is
about $T_c^*=0.168$). At such temperatures and densities the overall degree of polymerization $\Phi$ is always larger than 90\% (see Fig. \ref{Fig:PolymerizationKF}). The lines correspond to the same three functionals as shown in Fig. \ref{Fig:WDA_vs_FMT_hard_solute}. The quantity $\varepsilon_\text{sw}$ is chosen such that the peak height of $\rho(r)/\rho_b$ given by DFT for $\sigma < r < \sigma+\delta$ fits to simulations with a tolerance of a few per cent. \textcolor{black}{Note that we do not provide $\varepsilon_\text{sw}$ for each case, as it slightly varies for distinct density functionals and state points. Typically, we find $\varepsilon_\text{sw}/\varepsilon \approx 0.25 - 0.35$. }
 
Already at intermediate temperatures [(a), (b), and (e), (f)] where association between particles is not too strong, the mFMT performs clearly better than FMT in comparison to simulations. A correlation peak grows being located at $r \approx 1.75\sigma$, indicating an increasing tetrahedral ordering of the particles \cite{Horbach1999, Zaccarelli2007, Saika2013}. In contrast, the WDA functional (dashed lines in lower panel) completely fails to describe such a peak. For lower temperatures [(c), (d) and (g), (h)] the oscillations in simulations significantly become more pronounced; here, only the mFMT functional yields density profiles which are in adequate agreement with the simulations, although the correlations, and in particular the intensity of the second peak, are underestimated and slightly out of phase with simulations. 

 \begin{figure}[t] 
 	\centering
 	\includegraphics[width = 8cm]{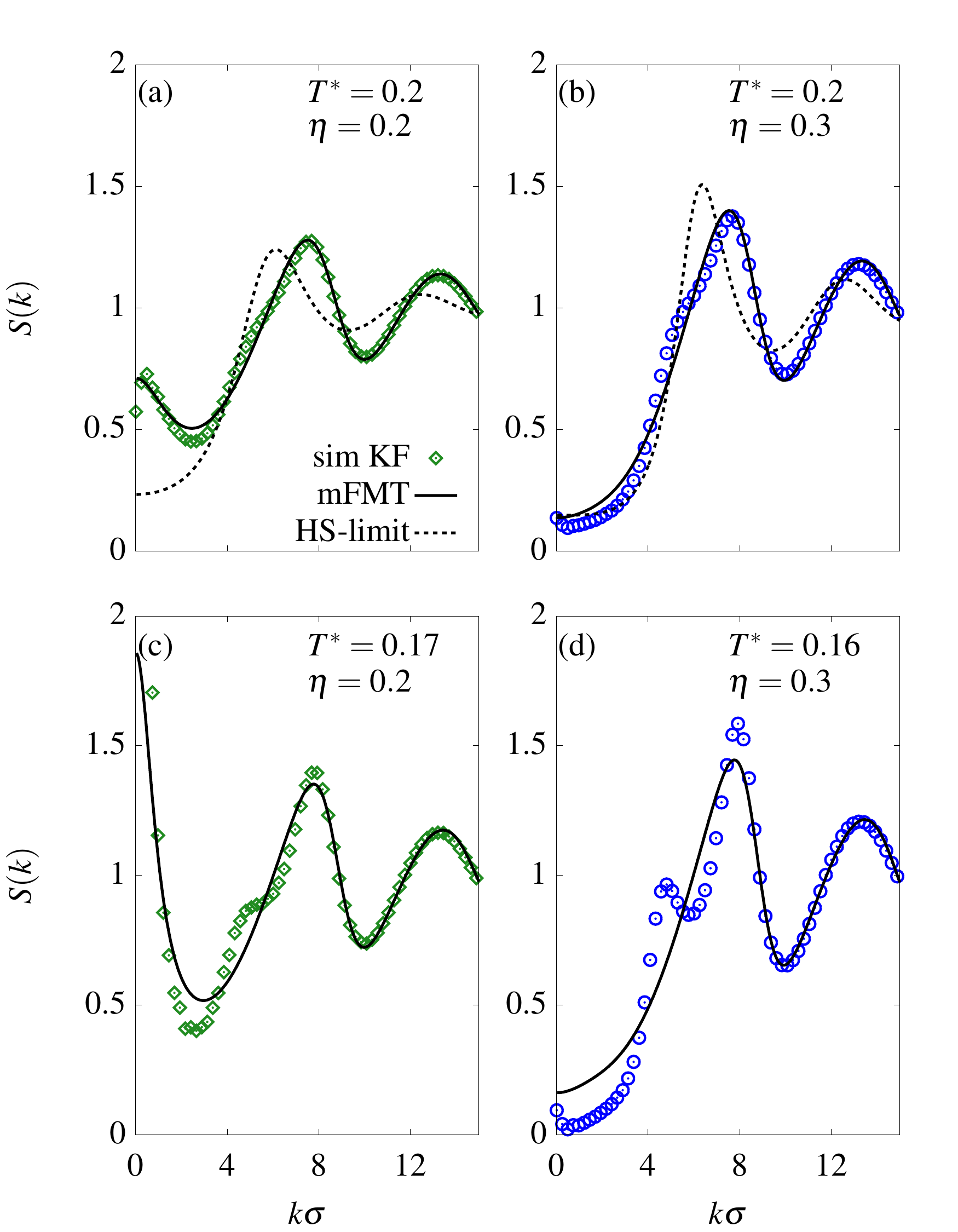}
 	\caption{Structure factors $S(k)$ obtained by numerically Fourier transforming theoretical density profiles (solid lines) and simulation results of patchy particles (symbols). The upper [lower] images show results for $T^* = 0.2$ [0.17 (c), 0.16 (d), respectively] with packing fractions $\eta = 0.2$ (left) and 0.3 (right). Dashed lines in (a) and (b) plot the corresponding Percus-Yevick structure factor for hard spheres.} \label{Fig:S_k_FMT_vs_sim}
 \end{figure}

 The overall good agreement between the latter and the theory is also reflected in momentum space when considering the static structure factor $S(k) = 1 + \rho_b \widehat{h}(k)$, where $\widehat{h}(k)$ is the three-dimensional Fourier transform of $g(r) - 1$. In Fig. \ref{Fig:S_k_FMT_vs_sim} we compare $S(k)$ extracted from simulations (symbols) and mFMT (solid lines) by numerically Fourier transforming the real-space data shown in Fig. \ref{Fig:FMT_vs_Patchy_eta02}. For higher temperatures $T^*=0.2$ and packing fractions $\eta=0.2$ (a) and 0.3 (b), we find excellent agreement between simulation and DFT, even in the low-$k$ region related to bulk thermodynamic properties of the system such as the isothermal compressibility. The dashed lines in (a) and (b) show the corresponding Percus-Yevick solution for the structure factor $S_\text{hs}(k)$ of hard spheres. For lower temperatures $T^* = 0.17$ (c) and 0.16 (d), however, deviations between DFT and simulations in the region $0 < k\sigma < 8$ become visible. In particular, for $\eta=0.3$ the static structure factors from simulations feature an additional peak at $k\sigma\approx 4.5$, which is absent in the theoretical calculations. Indeed, the presence of two peaks in $S(k)$ at $k\sigma\approx  4.5$ and $k\sigma\approx 8$ is a prominent signal of a system-spanning amorphous tetrahedral network \cite{Horbach1999, Zaccarelli2007, Saika2013}. While the shift of the monomer-monomer correlation peak from initially $k\sigma\approx 2\pi$ at high temperatures towards $k\sigma\approx 8$ at lower temperatures is also captured by the mFMT functional, we never observe a second peak at lower $k$, even for lower temperatures $T^* < 0.16$ than discussed here.
 These discrepancies are most likely due to approximations introduced by Wertheim's perturbation theory related to the absence of geometrical informations regarding the arrangement of bonding sites. This presumably makes it impossible to obtain an even better agreement in terms of bulk structural properties between a theory making use of TPT1 and simulations of patchy particles. 
  
 \begin{figure}[t] 
   	\centering
    \includegraphics[width = 8cm]{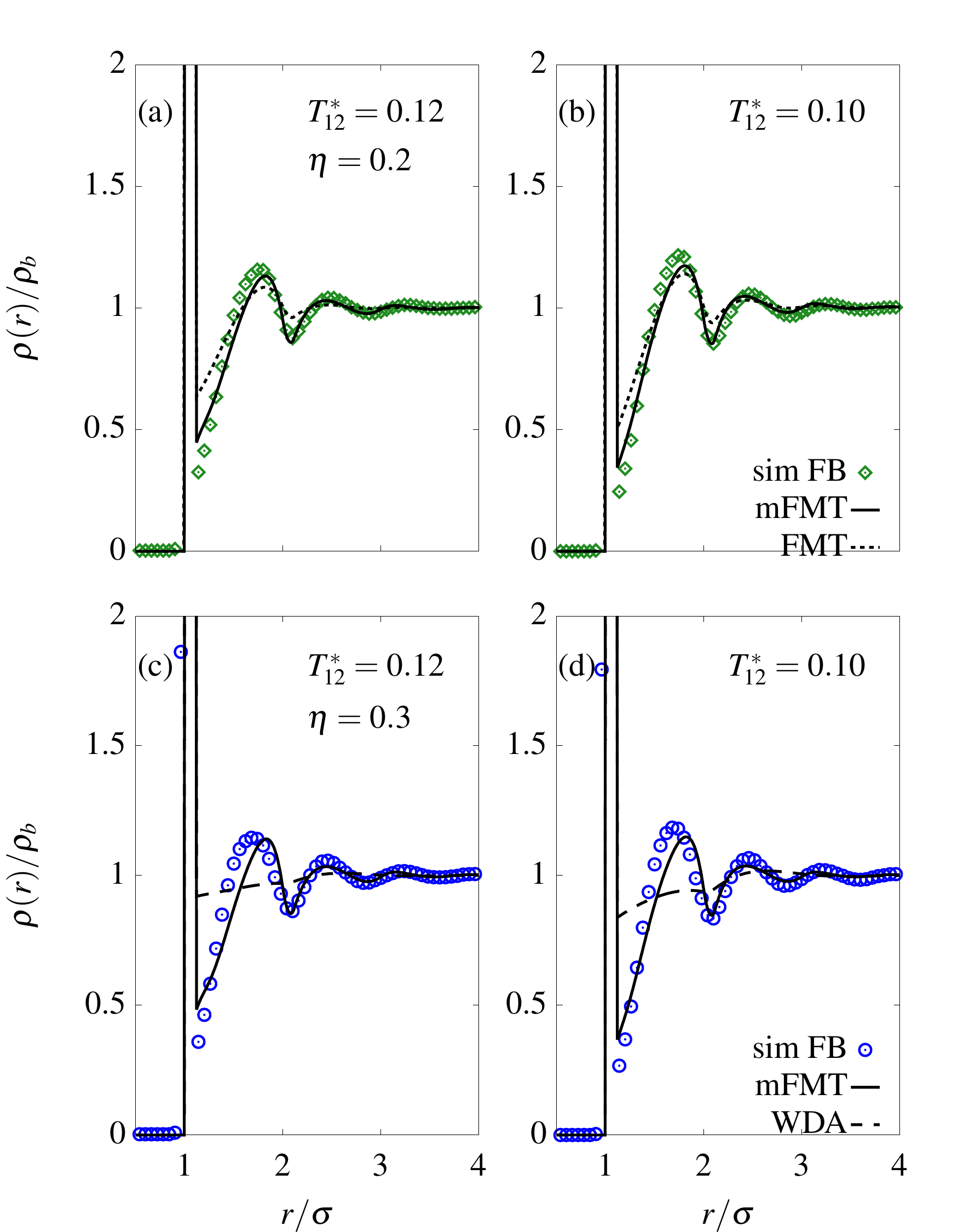}
   	\caption{Individual symbols: Simulation results for floating bonds at $\eta = 0.2$ for temperatures $T_{12}^* = 0.12$ (a) and 0.1 (b). (c) and (d) show the same at $\eta = 0.3$. Lines display corresponding temperature-rescaled results from DFT: mFMT (solid), FMT  (short-dashed), and WDA (long-dashed).} \label{Fig:FMT_vs_FB_eta02}
  \end{figure}      

 \subsection{Density functionals compared to simulations of the floating-bond model} \label{SubSec:CompFBDFT}

As already pointed out in Sec. \ref{SubSec:FloatingBonds}, the latter shortcomings are precisely the conceptual similarities between Wertheim's theory and the floating-bond model. Although the bond arrangement of the latter is not rigid, the system still prefers a tetrahedral ordering which can be concluded by considering the angular distribution $P(\theta)$ of triplets formed by particle-bond-particle-bond-particle structures \cite{Zaccarelli2007}. 
In Fig. \ref{Fig:FMT_vs_FB_eta02} we compare DFT results to the floating-bond model for $\eta = 0.2$ (upper panel) and $\eta = 0.3$ (lower panel), respectively, at temperatures $T^*_{12} = 0.12$ (left) and 0.1 (right). Recall that the critical point in simulations of floating bonds is approximately located at $(\eta_c,\,T_{12,c}^*) = (0.1,\, 0.105)$. The key is the same as in Fig. \ref{Fig:FMT_vs_Patchy_eta02}. The theoretical curves are rescaled according to the description of Sec. \ref{SubSec:FloatingBonds}; this means that e.g. a DFT result shown together with simulations for $T_{12}^* = 0.12$ corresponds to $T^* = 0.156$ in Wertheim's theory. The parameter $\varepsilon_\text{sw}$ is again chosen such that the overall agreement between theory and simulation is maximized within the region $\sigma < r < \sigma + \delta$. \textcolor{black}{Similar to the simulations of patchy particles, we find $\varepsilon_\text{sw}/\varepsilon_{12} \approx 0.25-0.35$.}  Importantly, we can conclude that the test-particle approach along with mFMT performs remarkably well on a quantitative level of agreement with the simulation results for both $\eta = 0.2$ and 0.3, which (as was the case in the previous section \ref{SubSec:CompSimDFT}) cannot be achieved using the original functional by Yu and Wu. In particular, the height of the correlation peak at $r \approx 1.75\sigma$ is accurately described by the theory, which stays clearly smaller than in case of patchy particles; the peak intensity does not  significantly increase further in simulations upon continuing cooling the system below $T^*_{12} = 0.1$. The only noticeable difference for $\eta=0.3$ is that the peak width is slightly broader than in theory. Note also that for the presented state points we have verified that the agreement between DFT and simulations is not qualitatively affected when employing hard potentials for the particle-particle and bond-particle interactions (i.e. hard-sphere and square-well potentials) instead of the here used \lq smooth\rq\,\,potentials (see Sec. \ref{SubSec:FloatingBonds} and Appendix \ref{SubSec:FloatingBondSimulationDetails}).

 \begin{figure}[t] 
 	\centering
 	\includegraphics[width = 8cm]{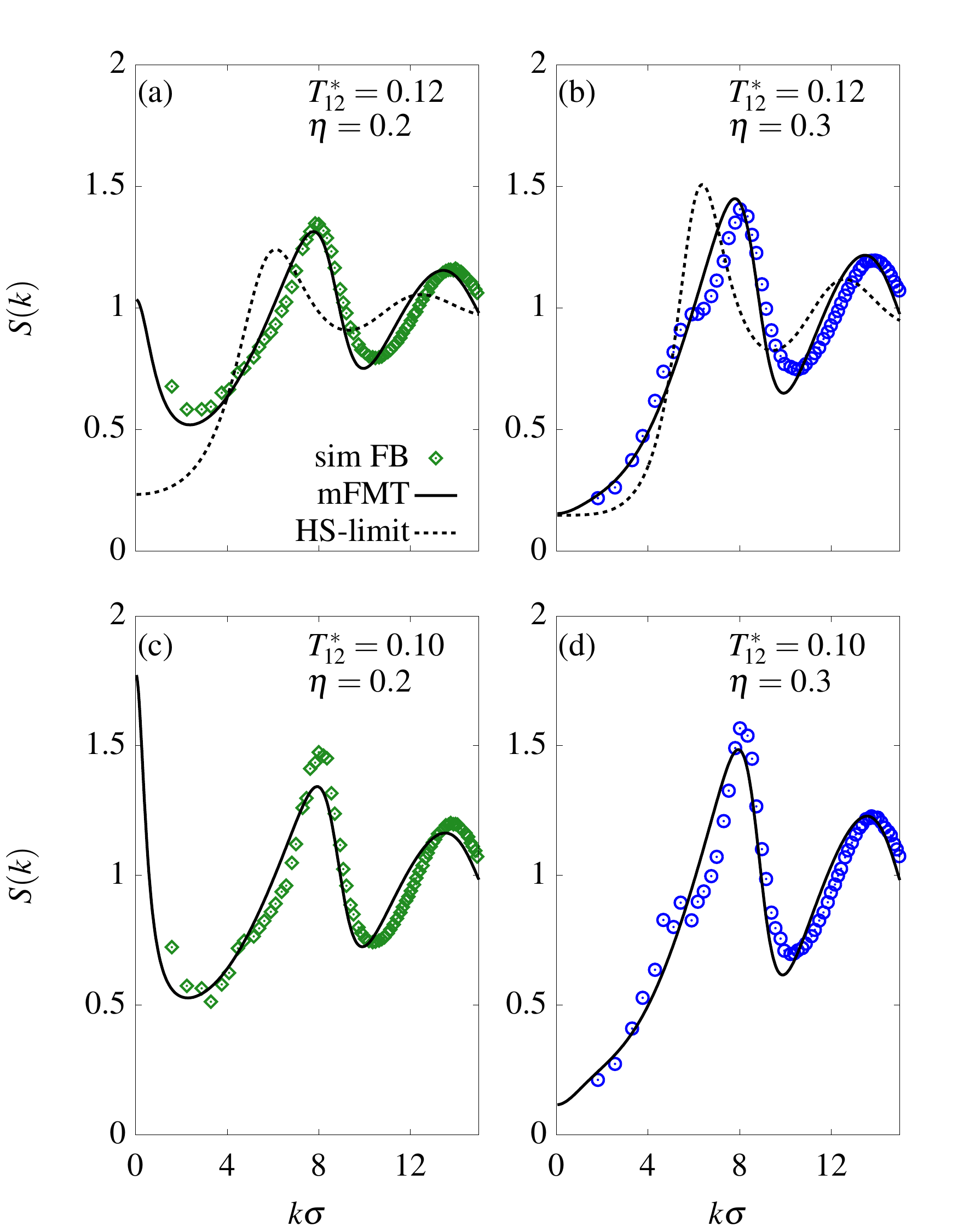}
 	\caption{Theoretical $S(k)$ (solid lines) compared to simulations of floating bonds (symbols). The upper [lower] images show results for $T^*_{12} = 0.12$ [0.10] with packing fractions $\eta = 0.2$ (left) and 0.3 (right). Dashed lines in (a) and (b) plot the corresponding Percus-Yevick structure factor for hard spheres.} \label{Fig:S_k_FMT_vs_FB}
 \end{figure} 
 
The static structure factor $S(k)$ for the floating-bond model also suggests that the tetrahedral ordering is weaker than in systems of patchy particles -- for the state points considered in Fig. \ref{Fig:FMT_vs_FB_eta02} and discussed above, we see from Fig. \ref{Fig:S_k_FMT_vs_FB} that the additional peak at $k\sigma \approx 4.5$ is absent or much weaker than found in simulations of patchy particles (cf. Fig. \ref{Fig:S_k_FMT_vs_sim}). 

\section{Discussion} \label{Sec:Discussion}

The fact that the Wertheim theory can be mapped nicely to the floating-bond model -- both in terms of thermodynamic properties such as the degree of polymerization $\Phi$, but also regarding the bulk structure -- may be traced back to conceptual analogies as already pointed out in Sec. \ref{SubSec:FloatingBonds}: In both models no statements are drawn regarding the distribution of bonding sites on the particle surface. Furthermore, screening effects that may become relevant due to volume exclusion effects for close-by sites are totally omitted in TPT1, and at least are minimized in the floating-bond model. In addition, the formation of ring-like structures is not taken into account by TPT1, and, in the floating-bond model there is only small signal at 60 $\deg$ in the angular distribution $P(\theta)$ of linked particle-triplets indicating three-particle rings \cite{Zaccarelli2007}. Note in particular that the situation is different to the study of Ref. \cite{Bianchi2008}, where it has been concluded that TPT1 does agree better with particles having an ordered and fixed patch arrangement than with random but fixed patch distributions. Indeed, an unordered (but rigid for each particle) arrangement of sites may favor formation of rings and give rise to non-negligible screening effects avoiding possible bonds due to steric incompatibilities at higher particle densities.

Our results impressively demonstrate that density functional theory making use of Wertheim's perturbation theory is capable of describing fundamental structural aspects prominent for fluids with directional interactions via an effective test-particle route. However, our findings also show that the quality of the results obtained from density functionals based on the bulk form of TPT1 seem rather sensitive on how precisely the generalization to the inhomogeneous fluid is achieved. In particular, for the FMT-based approaches this sensitivity can be traced back to the incorporation of the factor $\xi = 1 - \mathbf{n}_2\cdot\mathbf{n}_2/n_2^2$: When setting $\xi \equiv 1$, the FMT exhibits a very poor behavior, similar to the performance of the WDA functional. 
On the other hand, we have seen that replacing $\xi \rightarrow \xi^3$ yields accurate results in comparison to simulations  over a wide range of densities and temperatures (\textcolor{black}{we should also note that in planar geometries, such as introduced by a hard wall, the mFMT differs only marginally from the original FMT approach by Yu and Wu).}

\textcolor{black}{Furthermore, the remarkable performance obtained by minimization of the mFMT functional is rather interesting, as it turns out that the approach (as the WDA functional) does not respect the leading-order term of the exact virial expansion
\begin{align} \label{Eq:LowDensityLimit}
	\lim_{\rho\rightarrow 0} \beta F_\text{ex}[\rho] &= -\frac{1}{2} \iint \text{d}\mathbf{r}\,\text{d}\mathbf{r}'\, \rho(\mathbf{r})\rho(\mathbf{r}') \langle f(|\mathbf{r}-\mathbf{r}'|)\rangle \,,
\end{align}
where the (angular averaged) Mayer-$f$ function is given by
\begin{align} \label{Eq:LowDensityLimitc(r)}
 \langle f(r)\rangle
&= -\Theta(\sigma - r) + \mathcal{A} \left[\Theta(\sigma+\delta - r)-\Theta(r-\sigma)\right]\notag \\
&\equiv f_\text{hs}(r) + \langle f_\text{bond}(r)\rangle \,,
\end{align}
with $\mathcal{A} = M^2 (e^{\beta\varepsilon}-1) \sin^4(\theta_c/2)$. The latter
features a prominent peak for $\sigma < r < \sigma+\delta$ arising from nearest neighbor patch-patch interactions. 
By means of performing a functional Taylor expansion about $\rho = 0$ up to second order in density, one obtains the following low-density behavior of the FMT-type functionals (we consider only the bonding contribution, as the hard-sphere parts do respect Eq. \eqref{Eq:LowDensityLimit}) 
\begin{equation} \label{Eq:LowDensFMT}
\lim_{\rho\rightarrow 0}\beta F_\text{bond}[\rho] = -\frac{\mathcal{B}}{2}\int \text{d}\mathbf{r} \left(\left[n_0(\mathbf{r})\right]^2 - q \frac{\mathbf{n}_2(\mathbf{r})\cdot\mathbf{n}_2(\mathbf{r})}{\pi^2 \sigma^4}\right)\,.
\end{equation}
The factor $\mathcal{B}$ is given by $\mathcal{B} = M^2 (e^{\beta \varepsilon} - 1) v_b$, where the bonding volume $v_b$ is defined in Eq. \eqref{Eq:BondingVolume}. For completeness we also provide the low-density limit of the WDA functional,
\begin{equation} \label{Eq:LowDensWDA}
\lim_{\rho\rightarrow 0}\beta F_\text{bond}[\rho] = -\frac{\mathcal{B}}{2}\int \text{d}\mathbf{r}\, \rho(\mathbf{r}) \rho_0(\mathbf{r})\,.
\end{equation}
}
\textcolor{black}{Taking two functional derivatives one readily recognizes that Eqs. \eqref{Eq:LowDensFMT} and \eqref{Eq:LowDensWDA} do not yield $\langle f_\text{bond}(r)\rangle$ as given in Eq. \eqref{Eq:LowDensityLimitc(r)}. In the uniform fluid, however, both expressions are consistent with Eq. \eqref{Eq:LowDensityLimit} and are equivalent to the Wertheim free energy (Eq. \eqref{EqF_bond}) expanded up to second order in $\rho_b$,
 \begin{equation}
 	\lim_{\rho_b\rightarrow 0}\beta f_\text{bond} = - \frac{\mathcal{B}}{2} \rho_b^2 \,. 
 \end{equation}
 } 

\begin{figure}[t] 
	\centering
	\includegraphics[width = 8cm]{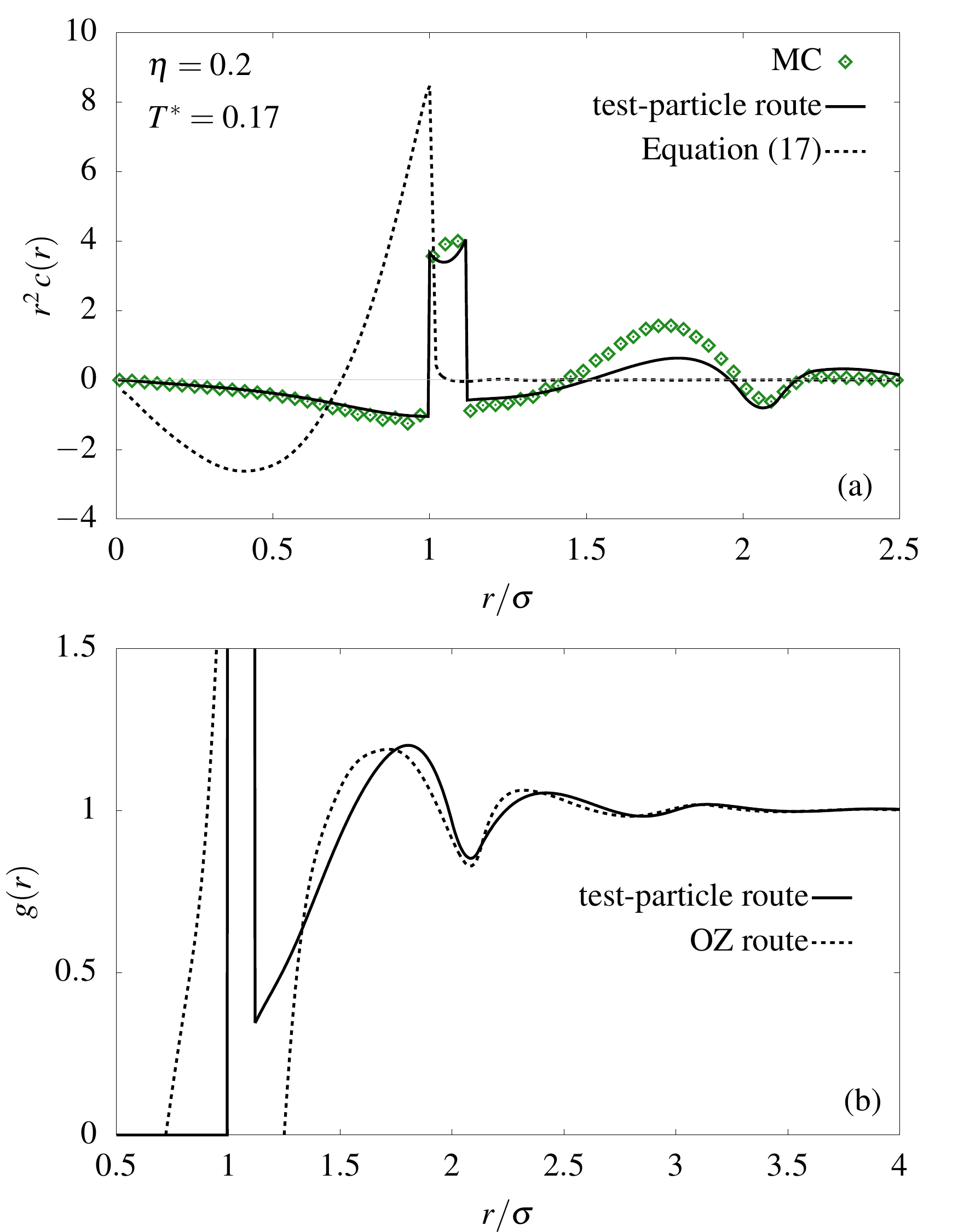}
	\caption{(a) Bulk pair direct correlation function $r^2\, c(r)$ obtained from simulations (symbols), functional differentiation according to Eq. \eqref{Eq:Defc(r)} (dashed line), and the test-particle route (solid line). The lines both show results from the mFMT functional. (b) Comparison of the $g(r)$ obtained from mFMT via the test-particle route (solid line) and from solving the Ornstein-Zernike equation \eqref{Eq:OZ} using the $c(r)$ obtained from Eq. \eqref{Eq:Defc(r)} (dashed line).  The state point in (a) and (b) is the same as in Fig. \ref{Fig:FMT_vs_Patchy_eta02} (d).} \label{Fig:cr}
\end{figure}

\textcolor{black}{The above shortcoming of the mFMT functional is also reflected when considering the full bulk pair direct correlation function $c(r)$ compared to simulations. Recall that it is defined as the negative second functional derivative of $F_\text{ex}[\rho]$ evaluated for the uniform fluid  (cf. Eq. \eqref{Eq:Defc(r)}). From simulations and integral equation theories it is known that $c(r)$ typically is negative inside the hard core due to repulsive interactions, and has positive contributions outside due to the presence of attractive forces \cite{Dijkstra2000}. 
Extracting a bulk pair direct correlation function $c_\text{sim}(r)$ from our simulations of patchy particles via the OZ relation, Eq. \eqref{Eq:OZ}, we observe a significant peak for $\sigma < r < \sigma + \delta$ in accordance with the Mayer-$f$ function (in the low-density limit it follows from Eq. \eqref{Eq:LowDensityLimit} that $c(r) = f(r)$) -- see Fig. \ref{Fig:cr} (a), where the symbols show $c_\text{sim}(r)$ for the same state point as in Fig. \ref{Fig:FMT_vs_Patchy_eta02} (d).} 
\textcolor{black}{In contrast, the $c(r)$ obtained from taking two functional derivatives of the mFMT functional exhibits a fundamentally distinct behavior (dashed line): It is zero for $r > \sigma$, but becomes positive inside the core. This can readily be understood since the FMT-type weight functions $\omega_\alpha$ are of range $\sigma/2$ and thus $c(r)$ is of range $\sigma$. As a result, the patchy attraction is completely mapped into the hard-core region.  \textcolor{black}{Note that there is no significant qualitative difference between the FMT and mFMT, albeit the former produces a less pronounced curve which goes in hand with the less structured density profiles in Fig. \ref{Fig:FMT_vs_Patchy_eta02}}. On the other hand, when extracting an effective $c_\text{eff}(r)$ from density profiles obtained via the test-particle route \textcolor{black}{and the mFMT approach}, it shows a similar level of good agreement with simulations as do the effective radial distribution functions which is impressively demonstrated by the solid curve in Fig. \ref{Fig:cr} (a). Remarkably, the $g(r)$ obtained from the OZ relation using $c(r)$ as generated by Eq. \eqref{Eq:Defc(r)} does not deviate too strongly from the $g(r)$ obtained via the test-particle route for $r \gtrsim 1.5\sigma$, see Fig. \ref{Fig:cr} (b). However, it does not exhibit the first-neighbor correlation peak for $\sigma < r < \sigma+\delta$ nor does it satisfy the core-condition $g(r) = 0$ for $r < \sigma$, which trivially is fulfilled in the test-particle situation.}

\textcolor{black}{Although the disagreement between the $c(r)$ obtained from directly differentiating the mFMT functional and via the test-particle route is striking, following Archer and Evans \cite{ArcherEvans2017} one should not judge a density functional based solely on its performance on the level of the (bulk) pair direct correlation function: for mean-field type DFTs, which are commonly employed for treating longer-ranged attractive spherically-symmetric pair interactions, it is well known that functional minimization according to Eq. \eqref{Eq:densityProfileEq} may lead to much more reliable results than taking two functional derivatives of $F_\text{ex}[\rho]$. We thus emphasize that the (m)FMT functional for patchy particles is not necessarily bound to yield unreliable results for the one-body density $\rho(\mathbf{r})$ in inhomogeneous situations.}
However, owing to the fact that the functional based on Tarazona's WDA in spherical geometries reveals to be unreliable in the present study, but in situations with planar geometry such as at a hard wall has shown reasonable agreement with simulations \cite{Segura1997}, indicate that this approach should indeed be applied with caution when aiming to investigate structural properties around, e.g., hard-sphere solutes. More generally, one should also bear in mind that there are a number of situations in which the orientation of particles may become relevant. For such cases, every density functional which does not respect the (average) particle orientations necessarily is inadequate and will give rise to deviations with computer simulations or experiments, as is demonstrated in Sec. \ref{SubSec:CompWDAFMT} and, e.g., in Ref. \cite{GnandelasHeras2012}.

 \section{Summary and Outlook} \label{Sec:Summary}
 
In this work, we employed classical density functional theory to address the bulk structure of associating fluids with a maximum number of bonded neighbors equal to four. To this end, we made use of density functionals incorporating Wertheim's first-order perturbation theory, which is known to yield reliable results for bulk properties such as the degree of polymerization (cf. Fig. \ref{Fig:PolymerizationKF}) and phase behavior, see, e.g., Refs. \cite{Jackson1988, BianchiLargo2006, Bianchi2007_JPCB, Bianchi2008, RussoTavares2011_PRL, Teixeira201716_review}. Three density functional approaches have been considered which commonly are used in the literature, one within the framework of Tarazona's weighted-density approximation \cite{Tarazona1990, Segura1997}. The second is based on Rosenfeld's fundamental measure theory \cite{Rosenfeld1989, Roth2010}, originally proposed by Yu and Wu \cite{YuWu2002}; the third functional that we considered in this work is a slight modified version of the latter. We calculated density profiles around a test particle interacting via an attractive square-well particles with its surrounding (patchy) neighbors, thereby obtaining effective radial distribution functions. In order to classify the results given by all the functionals, we first benchmarked the theoretical curves against results for the radial distribution function $g(r)$ obtained from computer simulations for  \lq real\rq\,\,patchy particles \cite{Rovigatti2018} having a fixed patch arrangement in tetrahedral order. It turned out that only with the modified FMT functional it was possible to match simulations satisfactorily, while the original FMT clearly underestimated correlations, and the WDA functional completely failed to account for essential structural properties. Nevertheless, upon lowering temperatures, even the extended FMT could not quantitatively describe the prominent correlation peak emerging in simulations, which is a structural indicator of increasing tetrahedral ordering of the particles. This, however, may be attributed to approximations introduced by Wertheim's theory itself rather than by (further approximative) density functionals: As part of the approximations, the theory does not provide informations about the geometry of the patches, which thus necessarily yields deviations between the former and simulations when e.g. comparing the bulk structure.

We have therefore also compared our theoretical results to simulations of a system which (to some extent) obeys conceptual analogies with Wertheim's first-order perturbation theory. The model generates directional interactions by means of a non-additive binary mixture of large and small particles, where only the small particles (so-called floating bonds) interact attractively with the large particles \cite{Zaccarelli2007}. The interaction potentials can be tuned such that one floating bond precisely connects  only two of the larger particles, with a maximum number of bonded small particles at one large particle equal to four. The ratio between floating bonds and particles can thereby be chosen such that in the ground state the system shows a tetrahedral structure. However, due to the flexibility of bonds, at finite temperatures the tetrahedral ordering naturally is not as strong as in patchy particles having a fixed tetrahedral symmetry. Indeed, by means of a proper temperature rescaling based on matching the degree of polymerization given by Wertheim's theory and the floating-bond model, we found that the structural properties generated by the modified FMT functional quantitatively fit nicely to radial distribution functions of the floating-bond model.

The present work poses directions for future work. For instance, besides structural indications regarding formation of particle networks upon increasing association, this typically is also reflected by dynamical quantities characterizing the particle mobilities such as the van Hove function $G(r,t)$ or, relatedly, the mean-square displacement \cite{Vargas2017}. In particular, dynamical density functional theory \cite{MarconiTarazona1999, ArcherEvans2004} has shown to provide a valuable tool to address structural relaxation not only of hard-body systems \cite{StopperHHGRoth2015JCP, StopperThorneywork2018}  but also for attractive systems \cite{StopperHHGRoth2016JPCM}. It would therefore be highly interesting to investigate whether, e.g., the modified FMT approach along with dynamic DFT is able to predict several features observed in simulations of patchy particles, such as a non-Gaussian behavior of the self-part of $G(r,t)$ or a prominent peak at $r = 0$ seen in the distinct part (characterizing the collective dynamic behavior of a system) related to a \lq hopping\rq-type motion of individual particles \cite{Vargas2017}. Dynamic DFT previously has shown to be capable of describing such phenomena in colloid-polymer mixtures where the polymers had a significant lower diffusivity than the colloids \cite{StopperHHGRoth2016JPCM}.

Second, it would be convenient to construct a functional which obeys the exact low-density limits Eqs. \eqref{Eq:LowDensityLimit} and \eqref{Eq:LowDensityLimitc(r)}, and shows a similar performance than the present FMT-based functional. An obvious choice is to follow Tarazona's mark I WDA \cite{Tarazona1984}, and introduce a suitable weighted density to be used in Eq. \eqref{EqF_bond} which by construction yields Eq. \eqref{Eq:LowDensityLimitc(r)}. \textcolor{black}{This can be achieved by defining a \lq bonding\rq\, weight function 
	\begin{equation}
	\omega_\text{bond}(r) \equiv \frac{\langle f_\text{bond}(r)\rangle}{\int \text{d}\mathbf{r}\, \langle f_\text{bond}(r)\rangle}\,,
	\end{equation}
	leading to a new weighted density $n_\text{bond}(\mathbf{r}) = \int\text{d}\mathbf{r}'\, \rho(\mathbf{r}') \omega_\text{bond}(|\mathbf{r}-\mathbf{r}'|)$.
	However, we found that this simple approach led to density profiles which are very poor compared to simulations, and thus more sophisticated approaches seem to be necessary for a consistent density functional theory for associating fluids.
	}  

\section*{Acknowledgments}
D. S. gratefully acknowledges funding by the Carl-Zeiss-Stiftung.

\appendix
 
 \section{Simulation details} \label{Sec:SimulationDetails}
 
 \subsection{Kern-Frenkel patchy particles} \label{SubSec:MCSimulationDetails}
 
All systems containing patchy particles interacting via the Kern-Frenkel potential \eqref{Eq:InteractionPotentialBonds} were simulated employing Monte-Carlo algorithms based on the open-source code \textit{PatchyParticles} introduced in Ref. \cite{Rovigatti2018}. We simulated several systems at temperatures $T^* = 0.5$--0.14, with volume fractions $\eta$ = 0.05, 0.2, and $\eta = 0.3$. Note that the critical point for the present model parameters is located at $\eta_c = 0.14$ and $T_{c}^* = 0.168$ \cite{Foffi2007}.  We employed the so-called AVB moves, that allow to significantly enhance sampling of the configurational phase space, in particular at lower densities \cite{Rovigatti2018}. The systems contained $N = 5000$ particles, and were equilibrated for $5\times 10^6$ Monte-Carlo steps, where the latter consists of $N$ individual attempts of performing an AVB move. Radial distribution functions were obtained by standard histogram-recording techniques, and production runs last for $10^5$--$10^6$ MC steps. Here, to minimize correlations between individual samples, only every $10$th to $100$th sample, depending on the temperature, contributed to a histogram. The density profiles around the hard and square-well tracer, respectively, were obtained  by hundred time longer production runs.

\subsection{Floating bonds} \label{SubSec:FloatingBondSimulationDetails}
We use the following continuous form for the interactions in simulations of the floating-bond model \cite{Bleibel2018}. The purely repulsive interactions among bonds and particles are given by the truncated and shifted Lennard-Jones potential
\begin{equation}\label{FBInteractionI}
\phi_{ii}(r) =
\begin{cases}
4\varepsilon_{ii} \left[\left(\frac{d_{ii}}{r}\right)^{12} - \left(\frac{d_{ii}}{r}\right)^6 + \frac{1}{4}\right]&;~~r < 2^{1/6}d_{ii} \\
~~~~~~~~~~~~~~~~~~0&;~~r \geq 2^{1/6}d_{ii}\,,
\end{cases}
\end{equation}
with $d_{ii} = 2^{-1/6}\sigma_{ii}$. The attraction and repulsion between bonds and particles is given by
\begin{equation}\label{FBInteractionII}
\phi_{12}(r) =
\begin{cases}
4\varepsilon_{12} \left[\left(\frac{d_{12}}{r}\right)^{12} - \left(\frac{d_{12}}{r}\right)^6 \right]&;~~r < 2^{1/6}d_{12} \\
~~-\frac{\varepsilon_{12}}{2}\left[1 + \cos\left(\frac{2\pi \widetilde{r}}{\lambda}\right)\right]~&;~~ 0 < \widetilde{r} < \frac{\lambda}{2} \\
~~~~~~~~~~~~~~~0&;~~ \widetilde{r} > \frac{\lambda}{2}\,,
\end{cases}
\end{equation}
in which $\widetilde{r} = r - 2^{1/6}d_{12}$, $\lambda = 0.134\, d_{12}$ and $d_{12} = 2^{-1/6}\sigma_{12}$. For simplicity, we set $\varepsilon_{11} = \varepsilon_{22} = \varepsilon_{12}$. Note also that the packing fraction $\eta$ neglects the volume of bonds and that we define a bond and a particle as bonded if $r_{12} < 0.58\sigma_{11}$.

Results regarding the floating-bond model have been obtained by conducting standard Brownian dynamics simulations for solving the position Langevin equation \cite{AllenTildesley1987}, for which the HOOMD-blue software package has been employed \cite{HOOMD_I,HOOMD_II}. The time step $\Delta t$ is determined by demanding an approximately constant particle displacement for each step and considered temperature. More explicitly, time is measured in units of Brownian time, i.e. $\Delta t = {\sigma^2_{11}}/{D_0} \Delta t_{\text{BD}} = {\sigma^2_{11}}{(\Gamma_1 T^*_{12} k_B)}^{-1} \Delta t_{\text{BD}}$, with the particle mobility $\Gamma_1$ and $k_B$ set to $1$.  The Brownian time is approximately the time a free particle needs to travel a distance equal to its own diameter, and we choose $\Delta t_\text{BD} = 6.66\times 10^{-7}$. 

The systems contain $N_1 = 1000$ particles and $N_2 = 2000$ bonds which each were equilibrated and measured for at least $7.5 \times 10^{7}$ time steps, whereas only every $1.25 \times 10^{4}$th step contributed to the measurement of an observable. For the data presented in Fig. 
\ref{Fig:FMT_vs_FB_eta02} and \ref{Fig:S_k_FMT_vs_FB} ten independent systems have been averaged.

\section{DFT implementations} \label{Sec:DFTImplementationDetails}
Due to symmetry, in this work the density profiles are radially symmetric, i.e. $\rho(\mathbf{r}) = \rho(r)$. We solved the effectively one-dimensional implicit equation \eqref{Eq:densityProfileEq} using a standard Picard algorithm on a discrete lattice with equidistant spacing of $\delta x = \sigma/100$. Iterations were terminated when the integrated relative change of the density profiles fell below $10^{-10}$. 

In spherical geometry, three dimensional Fourier transforms can be calculated straightforwardly using the Sinus transform,
\begin{equation}
\widehat{f}(k) = \frac{4\pi}{k} \int_{0}^{\infty} \text{d}r\, r f(r) \sin(kr)\,,
\end{equation}
and the inverse 
\begin{equation}
{f}(r) = \frac{4\pi}{(2\pi)^3 r} \int_{0}^{\infty} \text{d}k\, k \widehat{f}(k) \sin(kr)\,.
\end{equation}
Convolutions which frequently occur in classical DFT, efficiently can be calculated by above Fourier methods. In particular, the weight functions occurring both in WDA and FMT can be Fourier transformed analytically. These are given in the following.
\subsection{WDA weight functions in Fourier space} \label{SubSec:WDAweightFunctionsFourier}
For the WDA we have ($q = k\sigma$) \cite{Tarazona1990}:
\begin{align}
	\widehat{\omega}_0^\text{WDA}(k) &= \frac{3}{q^3}\left(\sin(q)-q\cos(q)\right)\,,\\
	\widehat{\omega}_1^\text{WDA}(k) &= \frac{\pi}{6}\sigma^3\frac{\widehat{\xi}(k) - 20\widehat{\omega}_0^\text{WDA}(k) - 10[\widehat{\omega}_0^\text{WDA}(k)]^2}{8[1 + \widehat{\omega}_0^\text{WDA}(k)]} \,,\\
	\widehat{\omega}_2^\text{WDA}(k) &= \frac{20\pi^2 \sigma^6}{q}\left(\frac{\cos(q)}{q} + \frac{4 + \cos(q)}{24q^3}-\frac{\sin(q)}{24q^2}\left[\frac{1}{q^2} + \frac{1}{2}\right]\right)\,,
\end{align}
with 
\begin{equation}
	\widehat{\xi}(k) = \frac{288}{q^6}\left[1 + q^2  - \left(1 + \frac{q^2}{2} + \frac{5q^4}{24}\right)\cos(q) - q\left(1 + \frac{q^2}{6}\right)\sin(q)\right].
\end{equation}

\subsection{FMT weight functions in Fourier space} \label{SubSec:FMTWeightFunctionsFourier}
The FMT-type weight functions in Fourier space are given by ($R = \sigma/2$)
\begin{align}
	\widehat{\omega}_3^\text{FMT}(k) &= \frac{4\pi}{k^3}\left(\sin(kR) - kR\cos(kR)\right)\,,\\
	\widehat{\omega}_2^\text{FMT}(k) &= \frac{4\pi}{k} R\sin(kR)\,, \\
	\widehat{\mathbf{\omega}}_2^\text{FMT}(\mathbf{k}) &= - i\mathbf{k}\, \widehat{\omega}_2^\text{FMT}(k)\,.		
\end{align}

\providecommand{\newblock}{}

\end{document}